\documentclass[aps,prd,superscriptaddress,preprintnumbers,amsmath,amssymb,latexsym,array,enumerate,letterpaper,floatfix,nofootinbib]{revtex4-2}

\usepackage{epsfig}
\usepackage[colorlinks=true,urlcolor=brown,linkcolor=blue,citecolor=magenta]{hyperref}
\usepackage{breakurl}
\usepackage{bm}
\usepackage[dvipsnames]{xcolor}
\usepackage{slashed}
\usepackage{cancel}
\usepackage{graphicx}
\usepackage{multirow}
\usepackage{comment}
\usepackage{epstopdf}
\usepackage{mathtools}
\usepackage{gensymb}
\usepackage{appendix}
\usepackage{url}
\usepackage[english]{babel}

\usepackage[skip=2pt plus1pt, indent=15pt]{parskip}

\makeatletter
\def\simgt{\mathrel{\lower2.5pt\vbox{\lineskip=0pt\baselineskip=0pt
           \hbox{$>$}\hbox{$\sim$}}}}
\def\simlt{\mathrel{\lower2.5pt\vbox{\lineskip=0pt\baselineskip=0pt
           \hbox{$<$}\hbox{$\sim$}}}}
\makeatother

\newcommand{\be}{\begin{equation}}
\newcommand{\ee}{\end{equation}}
\newcommand{\bea}{\begin{eqnarray}}
\newcommand{\eea}{\end{eqnarray}}
\newcommand{\beq}{\begin{eqnarray}}
\newcommand{\eeq}{\end{eqnarray}}
\newcommand{\Fig}[1]{Fig.~(\ref{#1})}
\newcommand{\Eq}[1]{Eq.~(\ref{#1})}
\newcommand{\Eqs}[2]{Eqs.~(\ref{#1}) and (\ref{#2})}
\newcommand{\Sec}[1]{Sec.~\ref{#1}}

\newcommand{\Subsec}[1]{Subsec.~\ref{#1}}

\newcommand{\eg}{\textit{e.g.}}
\newcommand{\ie}{\textit{i.e.}}

\newcommand{\neV}{\mathrm{neV}}

\newcommand{\MeV}{\mathrm{MeV}}
\newcommand{\GeV}{\mathrm{GeV}}
\newcommand{\TeV}{\mathrm{TeV}}

\newcommand{\hc}{\mathrm{h.c.}}

\newcommand{\MPl}{M_{\rm Pl}}
\newcommand{\mPl}{m_{\rm Pl}}

\newcommand{\infl}{\mathrm{inf}}
\newcommand{\rh}{\mathrm{rh}}
\newcommand{\osc}{\mathrm{osc}}
\newcommand{\oscu}{\mathrm{osc, 1}}
\newcommand{\oscd}{\mathrm{osc, 2}}
\newcommand{\mis}{\mathrm{mis}}

\newcommand{\finu}{\mathrm{end, 1}}

\newcommand{\iso}{\mathrm{iso}}

\newcommand{\on}{\mathrm{on}}
\newcommand{\off}{\mathrm{off}}

\newcommand{\ann}{\mathrm{ann}}

\newcommand{\fr}{\mathrm{fr}}
\newcommand{\pp}{\mathrm{pp}}
\newcommand{\str}{\mathrm{str}}
\newcommand{\mfp}{\mathrm{mfp}}

\newcommand{\QCD}{\mathrm{QCD}}
\newcommand{\PQ}{\mathrm{PQ}}
\newcommand{\NB}{\mathrm{NB}}

\newcommand{\PT}{\mathrm{PT}}

\newcommand{\MD}{\mathrm{MD}}

\newcommand{\MMo}{{\ensuremath{\mathrm{M \overline{M}}}}}
\newcommand{\MMoS}{{\ensuremath{\mathrm{M \overline{M}S}}}}

\newcommand{\tr}{\mathrm{tr}}

\newcommand{\LL}{\mathcal{L}}
\newcommand{\OO}{\mathcal{O}}
\newcommand{\VV}{\mathcal{V}}
\newcommand{\GG}{\mathcal{G}}
\newcommand{\TT}{\mathcal{T}}

\newcommand{\otheta}{\overline{\theta}}
\newcommand{\othetap}{\overline{\theta'}}

\newcommand{\SU}[1]{\mathrm{SU}(#1)}
\newcommand{\SO}[1]{\mathrm{SO}(#1)}
\newcommand{\Ua}{\mathrm{U}(1)}
\newcommand{\SUc}{{\SU{3}_c}}

\newcommand{\SUp}{{\SU{2}'}}
\newcommand{\Up}{{\Ua'}}

\newcommand{\abs}[1]{\ensuremath{\bl\vert{#1}\br\vert}}

\newcommand{\anti}[1]{\ensuremath{\overline{#1}}}
\newcommand{\vev}[1]{\ensuremath{\bl\langle {#1} \br\rangle}}

\newcommand{\bl}{\left}
\newcommand{\br}{\right}

\newcommand{\dd}{{\mathrm{d}}}
\newcommand{\Hinf}{H_{\text{inf}}}

\def\lsim{\mathrel{\rlap{\lower4pt\hbox{\hskip1pt$\sim$}}
     \raise1pt\hbox{$<$}}}         
\def\gsim{\mathrel{\rlap{\lower4pt\hbox{\hskip1pt$\sim$}}
     \raise1pt\hbox{$>$}}}         

\newcommand{\ignore}[1]{}

\usepackage{tikz,xcolor,hyperref}
\definecolor{lime}{HTML}{A6CE39}
\DeclareRobustCommand{\orcidicon}{%
	\begin{tikzpicture}
	\draw[lime, fill=lime] (0,0) 
	circle [radius=0.16] 
	node[white] {{\fontfamily{qag}\selectfont \tiny ID}};	\draw[white, fill=white] (-0.0625,0.095) 
	circle [radius=0.007];	\end{tikzpicture}
	\hspace{-2mm}}
\foreach \x in {A,...,Z}{%
	\expandafter\xdef\csname orcid\x\endcsname{\noexpand\href{https://orcid.org/\csname orcidauthor\x\endcsname}{\noexpand\orcidicon}}
}

\begin{document}

\title{Dynamical Axion Misalignment from the Witten Effect}

\author{Abhishek Banerjee\orcidA{}\,}
\email{abanerj4@umd.edu}
\affiliation{Maryland Center for Fundamental Physics, Department of Physics, University of Maryland, College Park, MD 20742, USA}

\author{Manuel A. Buen-Abad\orcidB{}\,}
\email{buenabad@umd.edu}
\affiliation{Maryland Center for Fundamental Physics, Department of Physics, University of Maryland, College Park, MD 20742, USA}
\affiliation{Department of Physics and Astronomy, Johns Hopkins University, Baltimore, MD 21218, USA}
\affiliation{Dual CP Institute of High Energy Physics, C.P. 28045, Colima, M\'{e}xico}

\begin{abstract}
    We propose a relaxation mechanism for the initial misalignment angle of the pre-inflationary QCD axion with a large decay constant.
    The proposal addresses the challenges posed to the axion dark matter scenario by an overabundance of axions overclosing the Universe, as well as by isocurvature constraints.
	Many state-of-the-art experiments are searching for QCD axion dark matter with a decay constant as large as $10^{16}\,\GeV$, motivating the need for a theoretical framework such as ours.
	In our model, hidden sector magnetic monopoles generated in the early Universe give the axion a large mass via the Witten effect, causing early oscillations that reduce the misalignment angle and axion abundance.
	As the hidden gauge symmetry breaks, its monopoles confine via cosmic strings, dissipating energy into the Standard Model and leading to monopole-antimonopole annihilation.
    This removes the monopole-induced mass, leaving only the standard QCD term.
	We consider the symmetry breaking pattern of $\SUp\to \Up\to 1$, leading to monopole and string formation respectively.
    We calculate the monopole abundance, their interactions with the axion field, and the necessary conditions for monopole-induced axion oscillations, while accounting for UV instanton effects.
    We present three model variations based on different symmetry breaking scales and show that they can accommodate an axion decay constant of up to $10^{16}\,\GeV$ with an inflationary scale of $10^{15}\,\GeV$.
	The required alignment between monopole-induced and QCD axion potentials is achieved through a modest Nelson-Barr mechanism, avoiding overclosure without anthropic reasoning.
\end{abstract}

\maketitle

\section{Introduction}

The discovery that charge-parity (CP) symmetry is not respected in Nature was one of the great surprises of 20th century particle physics.
In the Standard Model (SM), a CP-violating phase, present in the Cabibo-Kobayashi-Maskawa (CKM) matrix of the weak-interactions of the quark sector, is of order one \cite{Cabibbo:1963yz,Kobayashi:1973fv,ParticleDataGroup:2024cfk}.
Another CP-violating phase in the strong sector, often denoted by $\otheta$, can in principle also be present in the quantum chromodynamics (QCD) lagrangian.
However, constraints on the electric dipole moment of the neutron show that $\otheta$ must be smaller than $\approx 10^{-10}$ \cite{Baker:2006ts,Abel:2020pzs}.
The question of why the strong sector is CP-preserving while the weak sector violates it so badly is known as the ``strong CP problem'', and it is one of the most tantalizing hints that physics beyond the Standard Model (BSM) is needed.

Perhaps the most compelling solution to the strong CP problem involves a (pseudo-)Nambu-Goldstone boson, the ``axion'', which arises from the spontaneous symmetry breaking (SSB) of an anomalous $\Ua_{\rm PQ}$ global symmetry \cite{Peccei:1977ur,Peccei:1977hh,Weinberg:1977ma,Wilczek:1977pj,Zhitnitsky:1980tq,Dine:1981rt,Kim:1979if,Shifman:1979if}.
Non-perturbative QCD effects give this boson a potential (and consequently a mass) through a mixed anomaly between $\Ua_{\rm PQ}$ and the $\SU{3}_c$ gauge interactions of QCD.
When the axion field relaxes to its potential minimum, its vacuum expectation value (VEV) cancels the $\otheta$ phase, thereby explaining the aforementioned experimental results.
These non-perturbative effects relate the axion mass to its decay constant $f_a$, which is in turn closely related to the $\Ua_{\rm PQ}$-breaking scale.
This $f_a$ scale also controls the axion's coupling to the rest of the SM particle content, up to typically order-one model-dependent parameters.
Thus, the axion scenario is a highly predictive one, which makes it very attractive.

The elegant axion solution to the strong CP problem comes with an added bonus: axion particles can in fact have a cosmological abundance that explains the mystery of dark matter (DM), the ghostly substance making up about 80\% of all the matter in the Universe \cite{Planck:2018vyg}.
Indeed, thanks to the so-called axion misalignment mechanism \cite{Preskill:1982cy,Dine:1982ah,Abbott:1982af}, the axion field enters a phase of damped oscillations once its mass becomes comparable to the Hubble expansion rate of the Universe.
The amplitude of these oscillations dynamically evolves in such a way that the energy density stored in the axion field behaves just like DM should.
Thus, the cosmological abundance of the axion is tied to its mass (or $f_a$) and the initial value of the axion field, often parametrized in terms of a so-called initial ``misalignment angle'' $\theta_\mis$.

Depending on whether the $\Ua_{\rm PQ}$ symmetry was spontaneously broken before or after the end of inflation, respectively known as the pre- and post-inflationary axion scenarios, the axion can reproduce the DM abundance for various choices of $\theta_\mis$ and $f_a$.
Precise predictions of the axion energy density in both of these scenarios have been the subject of much work and continued debate; for a thorough review of axion cosmology see Ref.~\cite{OHare:2024nmr}.
Generically, the post-inflationary axion scenario has to contend with a variety of interesting yet potentially undesirable topological objects such as domain walls and strings, although this is model-dependent.
On the other hand, the pre-inflationary axion has no such issues, and the functional dependence of the axion abundance on $f_a$ and $\theta_\mis$ is more sharply defined than in the post-inflationary case.
However, the pre-inflationary scenario is instead limited by the fact that we do not know {\it a priori} the value of $\theta_\mis$ picked by the axion field in the inflationary patch that would become our Universe.
As it turns out, axion DM with a large $f_a$ requires small $\theta_\mis$, and vice versa.
In general, there are top-down reasons to expect $f_a$ to be very large; for example, string theory typically predicts a pre-inflationary axion with $f_a$ at the grand unified theory (GUT) scale \cite{Svrcek:2006yi,Reece:2024wrn}.
Furthermore, some experiments hoping to detect axion DM, such as DMRadio and CASPEr-electric, are {\it only} sensitive to relatively large $f_a$.
All of this seems to indicate that, given the typically large $f_a$ expected, a small $\theta_\mis$ is required for the pre-inflationary axion to be the DM; large $\theta_\mis$ values run in danger of overclosing the Universe.
Anthropic arguments can be made as to why this angle cannot be such that there is far too small or too large an axion abundance \cite{Linde:1987bx,Tegmark:2005dy}, although it is far from clear just how ironclad this line of reasoning really is.
This problem is compounded by the fact that the pre-inflationary axion will necessarily exhibit what are known as isocurvature fluctuations \cite{Linde:1984ti,Linde:1985yf,Seckel:1985tj,Turner:1990uz,Fox:2004kb}.
Indeed, since $\Ua_{\rm PQ}$ is broken during inflation, the light Nambu-Goldstone boson that is the axion will experience random fluctuations uncorrelated with those of the inflaton, which eventually gave rise to the anisotropies observed in the Universe.
Such axion isocurvature fluctuations are severely constrained by current cosmological observations \cite{Planck:2018jri}.
A sufficiently small inflationary scale, however, can evade such bounds; nevertheless, large-scale inflation is itself favored by many BSM frameworks.

Thus we arrive at an unavoidable three-way tension present in all pre-inflationary axion DM scenarios: a large-$f_a$ pre-inflationary QCD axion necessitates a tuned misalignment angle $\theta_\mis$, and it is furthermore incompatible with large-scale inflation \cite{Fox:2004kb}.
Recent years have seen the development of various theoretical mechanisms that can avoid these issues, such as entropy production \cite{Kawasaki:1995vt}, massive string moduli \cite{Banks:1996ea}, tachyonic instabilities \cite{Agrawal:2017eqm,Kitajima:2017peg}, or a QCD confinement scale that is inflation-dependent \cite{Co:2018phi}.
Of particular note are models where the misalignment angle enters a temporary phase of dynamical relaxation well before the onset of the more standard QCD axion oscillations.
Such a two-stage misalignment mechanism has the potential of addressing some of the problems outlined above.
Reference~\cite{Buen-Abad:2019uoc}, for example, considered a scenario where otherwise irrelevant contributions to the axion mass coming from small or UV instantons were much more important during inflation.
A larger axion mass at early times meant that the misalignment mechanism starts earlier, giving the axion angle extra time to relax, from the initial value $\theta_\infl$ picked by the axion at inflation, down to a much smaller $\theta_\mis$.
At later times these UV instantons cease to be important, freezing the axion angle to the $\theta_\mis$ value required by the DM abundance.
Similar models, where the axion potential evolves dynamically in such a way that the severe limitations imposed by cosmology on the axion DM scenario are relaxed, have been further explored in subsequent work \cite{DiLuzio:2021gos,Kitano:2021fdl,Berbig:2024ufe}.

The three-way tension of the pre-inflationary QCD axion scenario can be approached in yet a different way, involving the interplay of axions with yet another beautiful piece of field theory: magnetic monopoles.
Since they were first proposed by Dirac \cite{Dirac:1931kp}, and even more so after the discovery of the 't Hooft-Polyakov monopole solution four decades later \cite{tHooft:1974kcl,Polyakov:1974ek}, monopoles have been the object of intense study by theoretical physicists.
It can be shown, on mathematical grounds, that monopoles are in fact an unavoidable consequence of the unification of electromagnetism with other gauge forces into larger non-abelian gauge groups, such as in GUTs.
Monopoles could have been copiously produced in a phase transition in the early Universe, in a manner inconsistent with current observations \cite{Preskill:1979zi,Guth:1979bh,Einhorn:1980ym}; the search for a solution to this ``monopole problem'' was one of the driving factors behind the development of inflation.
Magnetic monopoles have all sorts of interesting field theoretical properties, one of the most striking being the Witten effect \cite{Witten:1979ey}.
According to this, a monopole in a background of non-zero CP phase will turn into a ``dyon'', an object that carries both magnetic and electric charges \cite{Julia:1975ff}.
Since the axion is essentially a dynamical CP phase, monopoles turn into dyons in the background of an axion field \cite{Fischler:1983sc}.
The potential energy stored in these axion-monopole interactions translates into an effective axion mass, proportional to the monopole abundance.
If this monopole-induced axion mass is large enough, the misalignment mechanism will cause the axion to enter its oscillatory phase much earlier than in the standard QCD axion picture, potentially relaxing the misalignment angle to sufficiently small values for axion DM to be viable.
While the tight bounds on the abundance of magnetic monopoles of the SM's electromagnetism preclude their Witten effect from dynamically relaxing the axion misalignment angle, magnetic monopoles of a hidden electromagnetic sector may do the job \cite{Fischler:1983sc}.
This has been recognized relatively recently in the literature, where hidden monopole-axion interactions have been leveraged to decrease the size of isocurvature \cite{Kawasaki:2015lpf,Nomura:2015xil,Kawasaki:2017xwt}.
However, these previous efforts either typically have a small $f_a$ and axion abundance, relying on the hidden monopoles to supply the rest of the dark matter \cite{Kawasaki:2015lpf,Kawasaki:2017xwt}; or invoke anthropic arguments to explain away the alignment coincidence between the minima of the monopole-induced and QCD axion potentials, necessary to prevent large-$f_a$ axions from overclosing the Universe \cite{Nomura:2015xil}.
Furthermore, neither a study of a plausible UV picture and its possibly negative impacts on the mechanism (such as irreducible UV instanton contributions to the axion mass that may spoil the axion solution to the strong CP problem), nor a systematic exploration of the parameter space, were undertaken in this pioneering research.

In this work we aim to do just that.
Based on an example UV model of the hidden sector, we are able to compute the cosmological abundance of the hidden magnetic monopoles and their interactions with the QCD axion, explain the origin of the requisite alignment of the two axion potentials, describe the necessary conditions for the dynamical relaxation of the axion misalignment angle through an early period of axion oscillations, carefully avoid the negative effects of UV instantons, and explore in detail the parameter space for which this mechanism works.

This paper is organized as follows.
In \Sec{sec:axion_dm} we review the basics of the pre-inflationary QCD axion DM scenario, and introduce the idea of the two-stage misalignment mechanism to relax the axion misalignment angle.
In \Sec{sec:model} we propose our benchmark model, outlining the conditions it must satisfy in order for the idea to work without invoking anthropics.
As stated in this introduction, we focus on the dynamical relaxation of the axion misalignment angle via the Witten effect of hidden monopoles.
We construct a plausible scenario which guarantees alignment of the monopole-induced and QCD axion potentials to the requisite level, detail the impact of UV instantons, and describe how and why the hidden monopoles must disappear.
Furthermore, we describe three different variants of our benchmark model, which correspond to different corners of its parameter space.
Section \ref{sec:mono} is devoted to computing the hidden monopole abundance, its interactions with the axion, the conditions under which this first stage of axion oscillations can occur, and the mechanism through which the monopoles can disappear.
Section \ref{sec:results} is concerned with the results of our study of the three versions of our model, showcasing the regions of their parameter space for which $f_a$ can be significantly larger than in the standard QCD misalignment mechanism.
Finally, we conclude in \Sec{sec:conclusions}.

\section{Axion Dark Matter}
\label{sec:axion_dm}

In the standard axion mechanism, an axion potential $\VV_\QCD$ arises from QCD non-perturbative effects. This potential is such that the VEV $\vev{a}$ of the axion field $a$ cancels out the physical, CP-violating QCD angle $\otheta$, thereby solving the strong CP problem \cite{Peccei:1977ur,Peccei:1977hh,Weinberg:1977ma,Wilczek:1977pj,Zhitnitsky:1980tq,Dine:1981rt,Kim:1979if,Shifman:1979if,GrillidiCortona:2015jxo,Hook:2018dlk}.
The potential $\VV_\QCD$ gives a mass to the axion, which at zero temperature is equal to \cite{GrillidiCortona:2015jxo,DiLuzio:2020wdo}
\beq\label{eq:ma0}
    m_{a,0} \equiv m_a \approx \frac{\sqrt{m_u m_d}}{m_u + m_d} \frac{m_\pi f_\pi}{f_a} \approx 0.57~\neV \ \bl( \frac{10^{16}~\GeV}{f_a} \br) \ ,
\eeq
where $f_a$ is the axion decay constant, $m_\pi$ and $f_\pi$ are the neutral pion's mass and decay constant, respectively; and $m_{u,d}$ are the up- and down-quark masses.
When the Universe was at a temperature above that of the QCD phase transition, $T_\QCD \approx 150~\MeV$, the axion mass from these non-perturbative effects was instead temperature-dependent:
\beq
    m_a(T > T_\QCD)^2 \approx m_a^2 \bl( T / T_\QCD \br)^{-n} \ ,
\eeq
with $n = 8$ providing a reasonably good approximation \cite{Borsanyi:2016ksw,DiLuzio:2020wdo}.

\subsection{The Misalignment Mechanism: Standard Lore}
\label{subsec:mis}

The misalignment mechanism \cite{Preskill:1982cy,Dine:1982ah,Abbott:1982af} allows for the axion field to make up part or all of the DM energy density in the Universe.
In the pre-inflationary axion scenario, with which we are concerned in this paper, the spatial average of the axion field ends up at a value $a_\infl$ in our patch of the Universe right at the end of inflation.
In terms of the axion angle $\theta \equiv (a - \vev{a}) / f_a$, we say that the axion is misaligned from its minimum by a value of $\theta_\infl$.%
\footnote{Throughout this paper, we use $\theta$ to the denote the {\it space-and-time-averaged amplitude} of the axion misalignment angle, \ie, the mean, over many oscillation periods, of the zero mode of the axion angle.
Note that the background axion energy density depends on the average of the {\it square} of the axion angle amplitude, while the isocurvature power depends on the mean square fluctuations of this angle.
These three distinct quantities are typically of the same order in the pre-inflationary scenario; therefore we do not distinguish between them.}
At early times the Hubble expansion rate of the Universe $H$ acts as a friction term on the axion field, keeping it frozen at a constant value.
Only at a time $t_\osc$, when the axion mass is comparable to the Hubble rate (more precisely, $m_a(T(t_\osc)) \approx 3 H(t_\osc)$), will the axion field start evolving in time by oscillating around its potential minimum.
In other words, $\theta_\mis \equiv \theta_\osc \approx \theta_\infl$ is the initial misalignment angle just before the axion starts to evolve.%
\footnote{Unless otherwise stated, we henceforth use $X_y$ to denote the value $X(t_y)$ of a dynamical quantity $X(t)$ at some time of interest $t_y$.}
If $t_\osc$ takes place during radiation domination, one can find that the temperature of the Universe at this time is $T_\osc = \bl( \sqrt{\frac{10}{\pi^2 g_*(T_\osc)}} m_a \mPl \br)^{\frac{2}{n+4}} T_\QCD^{\frac{n}{n+4}}$, which is close to $T_\QCD$ for large $n$.
Conservation of axion number in a comoving volume shows that after $t_\osc$ the axion field oscillates with an amplitude evolving like $\theta \sim T^{\frac{6-q}{4}}$, for an axion mass scaling as $m_a(T)^2 \sim T^q$.
Since after the QCD phase transition $q=0$, the amplitude evolves as $\theta \sim T^{3/2}$ and thus the axion energy density scales like $\rho_a = \frac{1}{2} m_a^2 f_a^2 \theta^2 \sim T^3$, as DM should.
An accurate, numerical computation of the axion abundance today yields \cite{GrillidiCortona:2015jxo,OHare:2024nmr}
\beq\label{eq:Omega_a}
    \Omega_a h^2 \approx 0.12~\bl( \frac{\theta_\mis}{2.15} \br)^2 \bl( \frac{f_a}{2 \times 10^{11}~\GeV} \br)^{1.16} \ .
\eeq
In this expression, $\theta_\mis \approx 2.15$ is an estimate of the root-mean-square of the initial misalignment angle over many inflationary domains, taking into account anharmonic corrections present in the QCD axion potential \cite{Fox:2004kb,GrillidiCortona:2015jxo,OHare:2024nmr}.
It is immediately clear what the problem is for the pre-inflationary axion DM: it requires inflation to set up an initial misalignment angle $\theta_\infl = \theta_\mis$ tuned to be very small for axion decay constants $f_a$ larger than $2 \times 10^{11}~\GeV$.
According to inflation, our Universe today arose from a single inflationary patch.
Thus, if the large-$f_a$ pre-inflationary axion scenario is true, our patch must be an atypical one: while $\theta_\mis \approx 2.15$ over many patches, in ours it accidentally has a much smaller value.
Such a coincidence is not inconceivable; in fact, anthropic arguments have been put forth in an attempt to explain why this must be the case \cite{Linde:1987bx,Tegmark:2005dy}.
Nevertheless, anthropic reasoning remains controversial, not in the least because these kind of arguments necessarily rely on a very limited analysis of the space of fundamental constants and model parameters.
Furthermore, a dynamical explanation for the smallness of parameters has significantly more explanatory power, and it is therefore to be preferred.

Regardless of the precise origin of axion DM, and of the mechanism that selects $\theta_\mis$ in our Universe, there are many good reasons to study large-$f_a$/small-$m_a$ axions.
For example, large $f_a$ values are favored by UV frameworks such as String Theory and small extra dimensions \cite{Svrcek:2006yi,Reece:2024wrn}.
In addition, experiments such as DM-Radio (with sensitivity in the interval $1.1 \times 10^{16}~\GeV \gtrsim f_a \gtrsim 7.1 \times 10^{12}~\GeV$) \cite{DMRadio:2022pkf} and CASPEr-electric (with a sensitivity above $f_a \gtrsim 1.9 \times 10^{15}~\GeV$) \cite{Budker:2013hfa,JacksonKimball:2017elr} are searching for QCD axion DM precisely in the large $f_a$ region \cite{AxionLimits}.
A recent axion DM search proposal relying on piezoelectric crystals may even probe $3 \times 10^{17}~\GeV \gtrsim f_a \gtrsim 10^{14}~\GeV$ \cite{Arvanitaki:2021wjk}.

Independently of whether it is DM or not, the pre-inflationary axion will experience spatial isocurvature fluctuations during inflation, with amplitudes given by $\delta a^\iso \approx \Hinf / 2 \pi$ \cite{Linde:1984ti,Linde:1985yf,Seckel:1985tj,Turner:1990uz,Fox:2004kb}.%
\footnote{Note that we have suppressed the dependence of these perturbations on the Fourier wavenumber $k$, \ie, $\delta a_k = \delta a$.
Recall that cosmological perturbations are defined with respect to the spatial average, \ie, $k=0$ zero-mode, of the quantity in question: $\delta a \equiv a_k - a$, where $a \equiv a_{k=0}$.}
Since $\delta \rho_a = m_a^2 \, a \, \delta a$, one can show that the temperature dependence of the axion perturbations for modes outside the horizon%
\footnote{All the modes relevant to current cosmological experiments were outside the horizon between the end of inflation and the onset of axion oscillations at $t_\osc$.}
is the same as that of its average field, namely $\delta \theta \equiv \delta a / f_a \sim T^{\frac{6-q}{4}}$.
This means that the ratio $\delta \theta / \theta$ does not evolve with time, and therefore the primordial power $A_\iso$ in isocurvature perturbations of the axion field at the onset of oscillations is trivially related to the same power at the time of inflation:
\beq
    A_\iso \approx \bl( \frac{\delta \rho_a^\iso}{\rho_a} \br)^2_\osc = \bl( \frac{2 \delta \theta^\iso}{\theta} \br)^2_\osc = \bl( \frac{2 \delta \theta^\iso}{\theta} \br)^2_\infl \approx \bl( \frac{\Hinf}{\pi \theta_\infl f_a} \br)^2 \ .
\eeq

For a flat isocurvature spectrum, the current 95\% C.L. bounds are $A_\iso < 0.038 A_s \approx 7.98 \times 10^{-11}$ \cite{Planck:2018jri}, for $\ln( 10^{10} A_s ) = 3.044$ the power in scalar adiabatic perturbations \cite{Planck:2018vyg}.
This means that $\delta \theta_\infl^\iso / \theta_\infl < 4.5 \times 10^{-6}$, or equivalently
\bea
    \Hinf & < & 1.2 \times 10^7~\GeV \ \bl( \frac{\theta_\infl}{2.15} \br) \bl( \frac{f_a}{2 \times 10^{11}~\GeV} \br) \ , \label{eq:iso_Hinfl}\\
    \Rightarrow \quad E_\infl & < & 7.2 \times 10^{12}~\GeV \ \sqrt{\frac{\theta_\infl}{2.15}} \sqrt{\frac{f_a}{2 \times 10^{11}~\GeV}} \ . \label{eq:iso_infl}
\eea
where for the last inequality we have used the inflationary energy scale $E_\infl \equiv (V_\infl)^{1/4} = (3 \mPl^2 H_\infl^2)^{1/4}$.
Enforcing the requirement that axions are DM, the standard cosmological evolution of the axion misalignment angle between inflation and $t_\osc$, namely $\theta_\mis \approx \theta_\infl$, implies low-scale inflation for large-$f_a$ axions (for example, $E_\infl < 7 \times 10^{13}~\GeV$ for $f_a = 10^{16}~\GeV$, since axion DM requires $\theta_\infl \approx \theta_\mis \approx 0.0041$).
Unless $f_a$ is outrageously large, \Eq{eq:iso_Hinfl} is a much stronger requirement than $\Hinf < 6.1 \times 10^{13}~\GeV$ (equivalently, $E_\infl < 1.6 \times 10^{16}~\GeV$), which is the 95\% C. L. constraint derived from the current bounds on the tensor-to-scalar ratio due to the non-observation of primordial gravitational waves by the Planck satellite in combination with the BICEP2/Keck Array \cite{Planck:2018jri}.

Equations (\ref{eq:Omega_a}) and (\ref{eq:iso_infl}) illustrate what, precisely, is the problem with large--$f_a$ axion DM in the standard cosmological picture: not only must the misalignment angle coming out of inflation be tuned so as not to overclose the Universe, but the scale of inflation itself must be very low in order to avoid dangerous levels of isocurvature.
Said differently: large $\Hinf$ is allowed for pre-inflationary axion DM but only for very large $f_a$ to compensate the smallness of $\theta_\infl \approx \theta_\mis$; yet in other words: misalignment angles of $\OO(1)$ are allowed, but at the cost of small $f_a$ and even smaller $\Hinf$.

\subsection{Two-Stage Dynamical Misalignment}
\label{subsec:two_mis}

\begin{figure}[t!]
\centering
\includegraphics[width=0.75\linewidth]{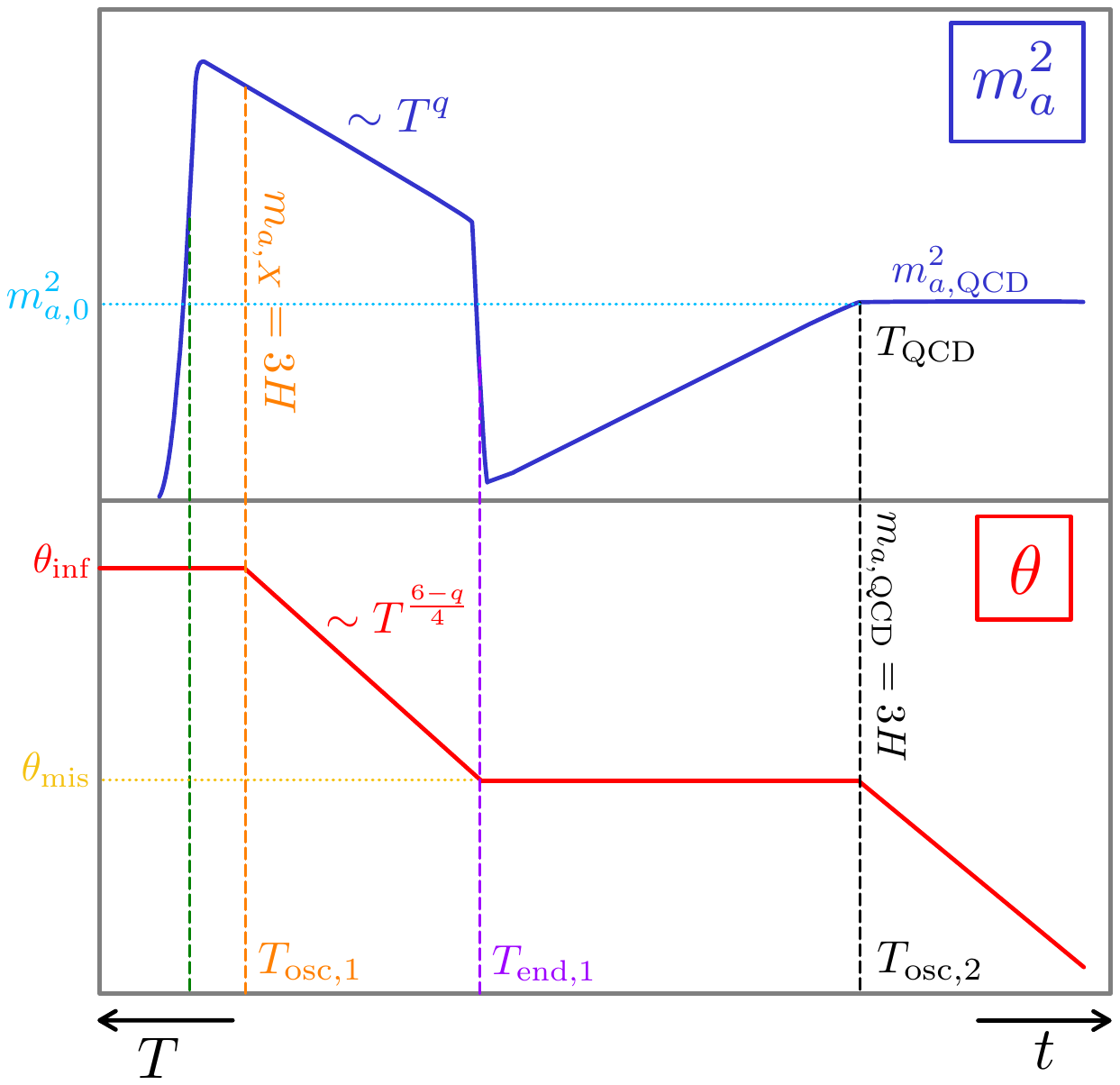}
\caption{Sketch of the evolution of the axion mass ($m_a^2$, blue curve, top panel) and the axion misalignment angle ($\theta$, red curve, bottom panel) required for the pre-inflationary axion to be the entirety of the dark matter.
At earlier times/large temperatures $\theta$ is fixed to $\theta_\infl$, the value picked right after inflation.
Some time later a new contribution $\VV_X(T)$ to the axion potential arises (dashed green line), providing it with a mass $m_{a,X}(T) \sim T^q$.
Eventually the oscillation condition $m_{a,X} = 3 H$ is satisfied when the Universe was at a temperature $T_\oscu$, and the misalignment angle relaxes from its initial value of $\theta_\oscu = \theta_\infl$, evolving as $\theta \sim T^{\frac{6-q}{4}}$ (dashed orange line).
This period of axion oscillations ends at $T_\finu$ (dashed purple line), when the potential $\VV_X(T)$ disappears; at this point $\theta$ reaches the value $\theta_\finu = \theta_\mis$ required by the standard misalignment mechanism (dotted yellow line).
Since the mass of the axion has dropped below $H$, the oscillation condition is no longer satisfied and the axion misalignment angle remains frozen at $\theta_\mis$.
The QCD axion potential $\VV_{a,\mathrm{QCD}}$ grows in importance until, at the time of the QCD phase transition at temperatures $T_{\rm QCD}$, it ceases to change with the temperature (dashed black line).
The axion mass takes its usual form, given by \Eq{eq:ma0} (dotted cyan line).
At around this time the axion mass once again satisfies the oscillation condition $m_{a,\mathrm{QCD}} = 3 H$, which allows the axion to enter a second era of oscillations with a new initial angle $\theta_\oscd = \theta_\mis$.
The axion field behaves like dark matter from this time onwards.
}
\label{fig:axion_mass}
\end{figure}

The astute reader will no doubt have noticed that the problems for the pre-inflationary scenario of axion DM stem from the assumption of a standard cosmological evolution for the misalignment angle between the end of inflation and the time when the axion field first begins to oscillate, around the QCD phase transition.
Instead, if the axion underwent an earlier period of evolution after inflation but before the QCD phase transition, then $\theta_\mis \equiv \theta_\osc \neq \theta_\infl$.
This would allow for a dynamical relaxation of the axion angle from a large $\theta_\infl$ to the small $\theta_\mis$ values necessary for the axion to reproduce the DM abundance with large $f_a$.
In addition, a larger $\theta_\infl$ makes it easier for the pre-inflationary axion to satisfy the bounds in \Eq{eq:iso_Hinfl} on isocurvature perturbations.
This earlier period of axion evolution can be induced by an additional but transitory contribution to the axion potential, which disappears before the QCD phase transition.
Thus, the axion potential would be
\beq
    \VV_a(T) = \VV_X(T) + \VV_\QCD(T) \ ,
\eeq
where $\VV_\QCD(T)$ is the QCD contribution to the axion potential, and $\VV_X(T)$ is the extra contribution, which first appears at some time $t_\on$ and goes away at $t_\off$.
The necessary earlier period of axion evolution occurs during the time interval $[ t_\oscu, t_\finu ]$ (with $t_\oscu \geq t_\on$ and $t_\finu \leq t_\off$) for which the axion mass $m_{a,X}(T)^2$ associated with $\VV_X$, obeys $m_{a,X}(T)^2 > 3 H^2(T), m_{a,\QCD}(T)^2$.
During this evolution $\theta$ is dynamically relaxed from its initial $\theta_\oscu \approx \theta_\infl$ value (typically $\OO(1)$) to the much smaller $\theta_\finu$.
At $t_\off$ $\VV_X$ goes away, and $\theta$ remains frozen at $\theta_\finu$.%
\footnote{Actually, there is a subtlety regarding the axion's oscillation frequency.
If the end of the $\VV_X$ contribution vanishes abruptly (\ie, in less than one Hubble time) at $T_\finu$ and catches the axion at a crest of its oscillation, $\theta_\finu$ will indeed remain frozen at this value from this time onwards.
If, however, $T_\finu$ catches the axion at a trough of the oscillation, the remnant velocity $\dot{\theta}_\finu \sim m_{a,X}(T_\finu) \theta_\finu$ will carry the axion away from the trough.
Since $m_{a,X}(T_\finu)$ drops quickly with $\VV_X$, it falls below the Hubble rate $H$ at $T \approx T_\finu$, and it's only then that the axion misalignment ceases to change.
This remnant velocity induces a change $\Delta \theta_\finu \sim (m_{a,X}(T_\finu) / H(T_\finu) ) \, \theta_\finu \sim \theta_\finu$ in the misalignment angle before it stops changing completely.
Therefore, regardless of where in the oscillation the axion is caught at $T_\finu$, the final amplitude of the axion misalignment is roughly $\theta_\finu$.
}
Only until a time $t_\oscd \geq t_\QCD$ shortly after the QCD phase transition does the second, standard period of axion oscillations begin, with an initial amplitude of $\theta_\mis \equiv \theta_\oscd \approx \theta_\finu$.
In \Fig{fig:axion_mass} we show a sketch of the evolution of the axion mass as a function of time, accompanied by its corresponding impact on the evolution of the amplitude of the axion angle.

As stated in the introduction, dynamical relaxation of the axion misalignment angle has received some attention in the last few years \cite{Buen-Abad:2019uoc,DiLuzio:2021gos,Kitano:2021fdl,Berbig:2024ufe}.
In this work we consider the scenario in which $\VV_X$ arises due to the interaction between the axion field and a population of monopoles and antimonopoles of a hidden sector (HS).
Through their coupling to the axions, the hidden monopoles gain an electric charge via Witten effect \cite{Witten:1979ey}, converting them into dyons and providing the axion with an additional mass term \cite{Fischler:1983sc}.
Previous literature has explored similar scenarios involving axions or axion-like particles (ALPs) coupled to hidden monopoles as a way to relax isocurvature bounds or allow for mixed DM made of axions or ALPs and the hidden monopoles themselves (see the original work in Refs.~\cite{Kawasaki:2015lpf,Nomura:2015xil,Kawasaki:2017xwt}, and more recent applications in Refs.~\cite{Sato:2018nqy,Nakagawa:2020zjr,Nakagawa:2021nme,Jeong:2023gjc}).
However, in this work we present for the first time how a pre-inflationary axion with a large $f_a$ can constitute the entirety of DM, without invoking neither anthropic reasoning or a tuning of the initial value $\theta_\infl$ of the axion angle \cite{Linde:1987bx,Tegmark:2005dy}, all while maintaining a consistent phenomenology and cosmology, and performing a careful study of the parameter space.

An important hurdle that dynamical misalignment models need to overcome is the fact that the minimum of $\VV_X$ need not be aligned with that of $\VV_\QCD$. Because of this, the evolution of the axion field in the earlier, monopole-induced period of oscillations will be centered around the VEV $\vev{a}_X$ that minimizes $\VV_X$, as opposed to around $\vev{a}$.
Thus, defining $\Delta \theta \equiv (\vev{a}_X - \vev{a}) / f_a$ as the offset between both potential minima, the quantity that represents the amplitude of the axion oscillations in this earlier evolution is the axion angle $(a - \vev{a}_X)/f_a = \theta - \Delta\theta$, and not $\theta$.
Assuming a power law behavior for the axion mass in this era, $m_{a,X}^2 \sim T^q$, we then have $\theta - \Delta\theta \sim T^{\frac{6-q}{4}}$, and thus the value of $\theta$ at $t_\finu$ is
\bea
    \theta_\finu & = & \bl( \frac{\theta_\finu - \Delta\theta}{\theta_\oscu - \Delta\theta} \br) (\theta_\oscu - \Delta\theta) + \Delta\theta \nonumber\\
    & = & \bl( \frac{T_\finu}{T_\oscu} \br)^{\frac{6-q}{4}} \ (\theta_\oscu - \Delta\theta) + \Delta\theta \ .
\eea
Since $\theta_\oscu \approx \theta_\infl$, and $\theta_\finu \approx \theta_\oscd \equiv \theta_\mis$, we have
\bea
    \theta_\mis  & = & D \ (\theta_\infl - \Delta\theta) + \Delta\theta \ , \label{eq:theta_mis}\\
    \text{where} \quad D & \equiv & \bl( \frac{T_\finu}{T_\oscu} \br)^{\frac{6-q}{4}} \label{eq:D_def}
\eea
defines the relaxation or dilution factor of the damped oscillations caused by the $\VV_X$ potential.

Thus, even if $D \ll 1$ and there were significant relaxation in the axion angle, we will never have $\theta_\mis \ll \theta_\infl \sim \OO(1)$ unless $\VV_X$ and $\VV_\QCD$ are aligned, \ie, $\Delta \theta \ll 1$.
It is therefore crucial for the minima of $\VV_\QCD$ and $\VV_X$ to coincide if we hope to have a dynamically small $\theta_\mis$ (as opposed to tuned), as required for the pre-inflationary, large-$f_a$ axion DM scenario to satisfy \Eqs{eq:Omega_a}{eq:iso_infl}.
Previous work \cite{Nomura:2015xil} using hidden monopoles to dynamically relax the axion misalignment and thus avoid isocurvature constraints simply assumed $\Delta\theta$ to be small, \eg~by invoking the anthropic principle.

Our challenge, then, is to build a model such that $\VV_X$ has a minimum almost exactly coincident with that of QCD; more precisely, we need $\Delta\theta/\theta_\infl \ll D \ll 1$.
If this is achieved, then \Eqs{eq:theta_mis}{eq:D_def} show us that the axion misalignment angle must be relaxed by a dilution factor of
\beq\label{eq:D_value}
    D = \frac{\theta_\mis}{\theta_\infl} \approx 0.0019 \, \bl( \frac{2.15}{\theta_\infl} \br) \sqrt{\frac{\Omega_a h^2}{0.12}} \bl( \frac{10^{16}~\GeV}{f_a} \br)^{0.58}
\eeq
in order for \Eq{eq:Omega_a} to be satisfied and the axion to be the entirety of the dark matter.

\section{The Model}
\label{sec:model}

Our goal in this section is to build a model such that ({\it i.}) there is an extra contribution to the axion potential, ({\it ii.}) whose minimum coincides with that of the QCD potential to a sufficient degree, and ({\it iii.}) eventually disappears before the QCD phase transition.
As stated in the introduction, the collective Witten effect caused by coupling the QCD axion to a population of monopoles of a hidden gauge group $\GG'$ in the early Universe can provide the additional potential required in condition {\it i.}, with the eventual disappearance of this monopole population being responsible for the temporary nature of this extra potential (condition {\it iii.}).
Condition {\it ii.} can only be satisfied if the $\otheta$ terms of both the QCD and $\GG'$ sectors are aligned.
Let us take each of these conditions in turn.

\subsection{Condition {\it i}: Monopoles and axion couplings}
\label{subsec:axion_mono_couplings}

In order for monopoles to exist, the gauge group $\GG'$ to which they belong must come from the breaking of a larger group $\overline{\GG'}$, in such a way that non-trivial topological gauge configurations are possible.
As such, there are many possible model realizations that have monopoles.
For concreteness we focus on a benchmark model where the gauge group is $\overline{\GG'} = \SUp$ and $g_2'$ is its gauge coupling.
$\SUp$ undergoes SSB when a scalar Higgs $\Sigma$ in the $\mathbf{3}$ (adjoint) irreducible representation (irrep) obtains a VEV, breaking the group down to $\GG' = \Up$.
This SSB Higgsing pattern results in $\mathbb{Z}$--monopoles, whose minimal charge is $h = 1/g_2'$ \cite{tHooft:1974kcl,Polyakov:1974ek}.%
\footnote{In reality, the result from this breaking is $\Up/\mathbb{Z}_2$, although the normal subgroup $\mathbb{Z}_2$ is often omitted in the literature.
For succinctness, we do likewise throughout the body of this paper.
Note that the second homotopy group of the vacuum manifold for this model is $\pi_2(\SUp/\Up/\mathbb{Z}_2) = \pi_1(\Up/\mathbb{Z}_2)/\pi_1(\SUp) = \mathbb{Z}/\mathbb{Z}_2$.
This means that monopole solutions exist, with the smallest magnetic charge equal to twice the minimal Dirac charge, \ie, $h = 1/g_2'$.
Of course, $\mathbb{Z}/\mathbb{Z}_2 \cong \mathbb{Z}$.
}
Models with other monopoles, \eg, $\mathbb{Z}_2$--monopoles of $\SO{3}$, would work just as well \cite{Garcia-Valdecasas:2024cqn}.
We postpone our discussion of the formation and abundance of $\Up$ monopoles to \Sec{sec:mono}.

In order for the required collective Witten effect to take place, it is necessary for the axion to couple to the larger $\SUp$ group so that, upon SSB, the $\Up$ inherits this coupling.
We rely on the PQ mechanism \cite{Peccei:1977ur,Peccei:1977hh} in its KSVZ version \cite{Kim:1979if,Shifman:1979if} to achieve this, with vector-like color quarks chirally charged under the $\Ua_\PQ$ symmetry.
Thus, the relevant terms of the lagrangian are
\beq
    \LL_\PQ \supset -y_Q \Phi^* Q Q^c + \hc - \lambda_\Phi \bl( \abs{\Phi}^2 - v_\PQ^2/2 \br)^2 \ ,
\eeq
where the $(\SUc, \SUp)_{\Ua_\PQ}$ charges of the fields are $Q \sim (\mathbf{3}, \mathbf{2})_{+1}$ and $Q^c \sim (\anti{\mathbf{3}}, \anti{\mathbf{2}})_0$ for the KSVZ quarks, and $\Phi \sim (\mathbf{1}, \mathbf{1})_{+1}$ for the PQ scalar.
For the pre-inflationary axion, the VEV $v_\PQ = \sqrt{2} \vev{\abs{\Phi}} > \Hinf$.
For our benchmark model with a single KSVZ vector-like pair, the anomaly coefficients are $N = 1$ and $N' = 3/2$, while $f_a \equiv v_\PQ/2N$.%
\footnote{The anomaly coefficients can be read from the result of performing a PQ transformation of $Q \to e^{i a / v_\PQ} Q$, which changes the lagrangian by the anomaly terms $\delta \LL = \frac{g_s^2}{16 \pi^2}\frac{a}{v_\PQ} \cdot 1 \cdot G_{\mu\nu}^a \tilde{G}^{a \mu\nu} + \frac{g_2'^2}{16 \pi^2}\frac{a}{v_\PQ} \cdot \frac{3}{2} \cdot W_{\mu\nu}^{\prime a} \tilde{W}^{\prime a \mu\nu}$.}
Note that the domain wall number is $N_{\rm dw} = 2 N = 2$, but since in the pre-inflationary scenario PQ symmetry is broken before inflation ends, the domain walls are inflated away.
The resulting Chern-Simons (CS) terms are
\beq\label{eq:axion_uv}
    \LL_a \supset \bl( \frac{a}{f_a} + \otheta \br) \bl( 
    \frac{\alpha_s}{8 \pi} \sum\limits_{a=1}^{8} G^a_{\mu\nu} \tilde{G}^{a \mu \nu} \br) + \bl( \frac{a}{f_a} \frac{N'}{N} + \othetap \br) \bl( \frac{\alpha_2'}{8 \pi} \sum\limits_{i=1}^{3} W^{i \prime}_{\mu\nu} \tilde{W}^{i \prime \mu\nu} \br) \ ,
\eeq
where $\alpha_s$ and $G^a$ denote the QCD fine structure constant and gluon stress tensor of $\SUc$, respectively, while $\alpha_2'$ (defined as $\alpha_2' \equiv g_2'^2 / 4\pi$) and $W'$ are the $\SUp$ fine structure constant and stress tensor.
Here $\otheta$ and $\othetap$ denote the $\SUc$ and $\SUp$ CP phases.
The tilde denotes $\tilde{X}^{\mu\nu} \equiv \frac{1}{2} \epsilon^{\mu\nu\rho\sigma} X_{\rho\sigma}$.

When $\Sigma$ gets a VEV and Higgses $\SUp \to \Up$, the hidden photon is $A' \equiv W^{\prime 3}$, and $e' \equiv g_2'/2$ the hidden electric gauge coupling (so that the smallest hidden electric charge is $1$).
Defining $\alpha' \equiv e^{\prime 2} / 4 \pi$ and $F'$ as the hidden electromagnetic tensor, we find the low-energy effective axion lagrangian:
\beq\label{eq:axion_eft}
    \LL_{a} \supset \bl( \frac{a}{f_a} + \otheta \br) \bl( 
    \frac{\alpha_s}{8 \pi} \sum\limits_{a=1}^{8} G^a_{\mu\nu} \tilde{G}^{a \mu \nu} \br) + \bl( \frac{a}{f_a} \frac{E'}{N} + 4 \othetap \br) \bl( \frac{\alpha'}{8 \pi} F'_{\mu\nu} \tilde{F}^{\prime \mu\nu} \br) \ ,
\eeq
where $E' \equiv N' \cdot (g_2'/e')^2 = 6$ is the resulting hidden electromagnetic anomaly.

The second term in \Eq{eq:axion_eft} is the one responsible for the Witten effect around the hidden monopoles.
Indeed, in the presence of a background $a$ or $\othetap$, it gives the monopoles an electric field, turning them into dyons \cite{Schwinger:1969ib,Julia:1975ff,Witten:1979ey}.
Because electric fields carry electrostatic energy, a dyon has a larger mass than a monopole.
Thus, in order to minimize the energy of the system, the dynamical axion field will settle at the minimum $\vev{a}_M/f_a = - 4 N \othetap / E' = - 2 \othetap / 3$ at the location of the monopole.
In terms of the axion fluctuations around this minimum, $\varpi \equiv (a - \vev{a}_M)/f_a$, the potential energy from the axion-monopole coupling, for a monopole of magnetic charge $h$, is given by \cite{Fischler:1983sc}
\bea
    V_0 & \approx & \beta f_a \varpi^2 = \bl( \frac{2 \beta}{f_a} \br) \frac{1}{2} \bl( a - \vev{a}_M \br)^2 \ , \label{eq:axion_dyon_potential} \\
    \text{with} \quad \beta & \equiv & \frac{e'^2 \, \bl( E'/N \br)^2 \, \bl( 2 e' h \br)^2}{128 \pi^3 r_a f_a} = \frac{\alpha'}{32 \pi^2} \frac{\bl( E'/N \br)^2 \, \bl( 2 e' h \br)^2}{r_a f_a} \ , \label{eq:beta_axion}
\eea
where $r_a = \max\bl[ r_0, r_c, m_f^{-1} \br]$ is the characteristic length of the axion-monopole system. The length scale $r_0$ is defined as $r_0 \equiv (E'/N) (2e' h) \, e' / (16\pi^2 f_a)$ and is important when the monopole appears point-like; $r_c \approx 1 /(g_2' v_\Sigma) = 1 /(4 \alpha' m_M)$ is the core size of a monopole of mass $m_M \approx 4 \pi v_\Sigma / g_2' = \sqrt{\pi/\alpha'} \, v_\Sigma$, and $m_f^{-1}$ is the Compton wavelength of the lightest charged fermion.
In our benchmark model we have $r_a = \max\bl[ r_c, m_f^{-1} \br]$, since we typically have $r_0 / r_c \approx \bl( E'/N \br) \bl( 2 e' h \br) \bl( \alpha'/2 \pi \br) \bl( v_\Sigma / f_a \br) < 1$ for most of our parameter space.
From now on we consider only monopoles with the minimal Dirac charge $h = 1/g_2' = 1/(2e')$, and thus take $2e'h = 1$ in $\beta$.

\subsection{Condition {\it ii}: Approximate alignment of axion potentials}
\label{subsec:nelson_barr}

Since the minimum of the axion monopole potential is $\vev{a}_M / f_a = - 2 \othetap / 3$, while the minimum of the axion QCD potential is $\vev{a} = - \otheta$, the alignment between both potentials can be stated as $\Delta\theta = \otheta - 2\othetap/3 \approx 0$, or $\othetap \approx 3 \otheta/2$.
As stated in the previous section, the level at which these two angles must be equal depends on the amount of dilution $D$ necessary for the axion abundance to be that of DM.
The simplest way to satisfy this requirement is by assuming a CP-symmetric Universe, where spontaneous CP symmetry breaking \`{a} la Nelson-Barr (NB) \cite{Nelson:1983zb,Barr:1984fh,Barr:1984qx} is ultimately responsible for the weak CP phase in the CKM matrix of the quark sector.
This ensures $\othetap = 0$ (if there is no hidden quark Yukawa sector with its own phases) and $\otheta \ll 1$ if the usual loop-induced NB QCD CP-phase is sufficiently small.
Note that in our scenario the purpose of the NB mechanism is {\it not} to solve the strong CP problem; that is the role of the QCD axion.
Instead, all NB needs to do is ensure that the QCD and monopole potentials have their minima aligned to within $\Delta \theta \ll D \theta_\infl \approx 0.0041 \times (10^{16}~\GeV / f_a)^{0.58}$ (see \Eq{eq:D_value}); a much milder requirement than the usual $\otheta < 10^{-10}$.

The NB sector can be similar to that of the Brento-Branco-Parada (BBP) model \cite{Bento:1991ez}:
\beq
    \LL_\NB \supset -\VV_\eta - m_Q \, Q' Q^{\prime c} - (a_i \eta + a_i' \eta^*) \, d_i^c Q' + \hc \ ,
\eeq
where $\eta$ is the NB complex scalar whose VEV $\vev{\eta} \equiv v_\NB \, e^{i \delta} / \sqrt{2}$ minimizes the potential $\VV_\eta$ and spontaneously breaks CP symmetry \cite{Bento:1990wv}, $a_i \neq a_i'$ are two complex parameters, $d_i^c$ are the down-type SM quarks, $i \in \{ d, s, b \}$ are the three down-type flavors, and $Q'$ and $Q^{\prime c}$ are the additional $\mp 1/3$ charge, $\SU{2}_w$-singlet, NB quark--anti-quark pair.
For the NB mechanism to work, we must have $m \lesssim \abs{B_i}$, where $B_i \equiv a_i \vev{\eta} + a_i' \vev{\eta^*}$, and the $\eta Q' Q^{\prime c}$ and $H q_L Q^{\prime c}$ terms must be forbidden.
The latter can be ensured if $Q'$, $Q^{\prime c}$, and $\eta$ are charged under a discrete $\mathbb{Z}_2$ global symmetry.
We could go even further and let the PQ symmetry take care of this: if $Q^{\prime c}$ is charged under $\Ua_\PQ$, the NB quark mass term can arise from $y_Q' \Phi Q' Q^{\prime c}$, and the dangerous operators cannot be constructed.
In this case, the requirement $m \lesssim \abs{B_i}$ means that $y_Q' v_\PQ \lesssim \vert a_i^{(\prime)} \vert v_\NB$.
Barring an {\it ad hoc} large hierarchy between $y_Q'$ and $\vert a_i^{(\prime)} \vert$, this means that the scales of spontaneous CP- and PQ-breaking by the $\eta$ and $\Phi$ scalars are roughly similar, which suggests a common origin.
We leave the model-building of such an intriguing possibility to future work.
In this joint PQ--NB scenario, since $v_\PQ > \Hinf$, we need $v_\NB > \Hinf$ as well.

Additional terms such as $c_1 (\eta^2 + \eta^{*2}) \abs{H}^2 + c_2 \abs{\eta}^2 \abs{H}^2$ will contribute to $\otheta$ at one loop, leading to the rough scaling (with $a_i^{(\prime)} \sim a$ and $c_{1,2} \sim c$) \cite{Bento:1991ez,Dine:2015jga}:
\beq\label{eq:delta_theta}
    \delta \otheta \sim \frac{a^2 c}{16 \pi^2} \ .
\eeq
Therefore, $\otheta = \delta\otheta \neq 0$, and, since $\otheta' = 0$, we have $\Delta\theta = \delta\otheta$ as well.
This is safely below the required value for modestly small $a_i^{(\prime)}$ and $c_{1,2}$ coefficients.
We have thus demonstrated that a low-quality NB mechanism is sufficient to align the minima of the QCD- and monopole-induced axion potentials, so as to allow for the dynamical relaxation of the axion misalignment angle to work.
To reiterate, our solution to the strong CP problem is not of the NB kind, but the axion.
Nelson-Barr spontaneous CP symmetry breaking simply ensures that the monopole and the QCD sectors have their CP phases aligned to the requisite level.

The above discussion assumes that at low energies the dominant contribution to the axion potential comes from QCD.
If additional $\SUp$ contributions coming from the second term of \Eq{eq:axion_uv} were to exist and dominate the potential, then the axion VEV would settle at $\othetap = 0$ instead of at $\otheta = \delta \otheta$.
This would mean that there is a remaining $\otheta$ much larger than the experimental bound of $\otheta < 10^{-10}$, thus spoiling the axion solution to the strong CP problem.
As it turns out, such $\SUp$ contributions do exist, and arise from UV or small instantons evaluated at the mass scale $v_\Sigma$ associated with the SSB of $\SUp$.
In the dilute instanton gas approximation \cite{tHooft:1976snw,Callan:1977gz,Andrei:1978xg,Csaki:2023ziz}, the mass contribution from these UV instantons can be estimated as
\bea
    m_{a,\mathrm{UV}}^2 f_a^2 & \approx & 
    D_{\rm ins}(v_\Sigma) K v_\Sigma^4 \ , \label{eq:uv_inst} \\
    \text{with} \quad D_{\rm ins}(\mu) & \approx & 0.1 \bl( \frac{2 \pi}{\alpha_2'(\mu)} \br)^4 \, e^{-\frac{2 \pi}{\alpha_2'}(\mu)} \ ,\\
    \text{and} \quad K & \equiv & \prod\limits_i^{N_f} \frac{m_i}{v_\Sigma} \ .
\eea
Here $D_{\rm ins}$ is the instanton density, and $K$ is the chiral suppression from $N_f$ vector-like HS fermions, of mass $m_i$, in the $\mathbf{2}$ irrep of $\SUp$.

\begin{figure}[t!]
\centering
\includegraphics[width=0.75\linewidth]{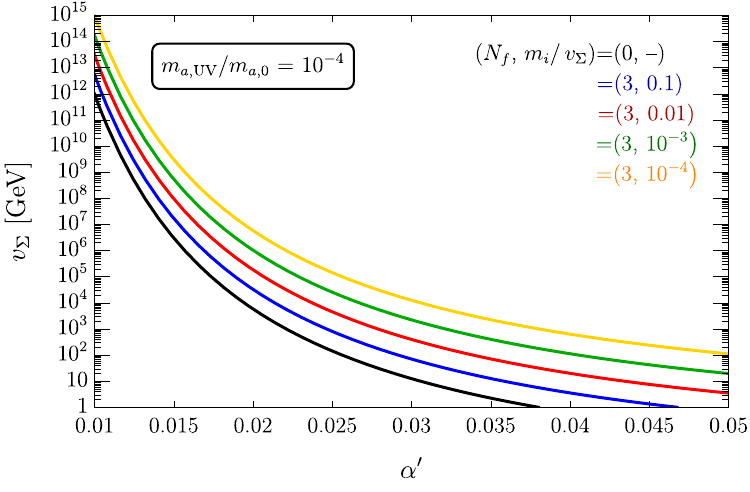}
\caption{The largest allowed value of the $\SUp$ SSB scale $v_\Sigma$ as a function of $\alpha' = \alpha_2'/4$, obtained by demanding that the contribution from UV instantons to the axion mass, given by \Eq{eq:uv_inst}, be at most $10^{-4}$ of the QCD value, \Eq{eq:ma0}, in order to protect the solution to the strong CP problem.
The black line assumes there are no extra vector-like fermion doublets with masses $m_i$ below $v_\Sigma$.
The blue, red, green, and yellow lines correspond to $N_f = 3$ extra vector-like fermion doublets with degenerate masses such that $m_i / v_\Sigma = 0.1, \, 0.01, \, 10^{-3}, \text{ and } 10^{-4}$, respectively.}
\label{fig:uv_instanton}
\end{figure}

Demanding that these UV instantons do not spoil the QCD axion VEV to the level of $10^{-10}$ is equivalent to requiring $m_{a,\mathrm{UV}} / m_{a,0} \lesssim \sqrt{10^{-10} / \delta\otheta} = 10^{-4} \sqrt{10^{-2} / \delta\otheta}$.
This condition allows us to find the largest possible value of $v_\Sigma$ for a choice of $\alpha_2' = 4 \alpha'$, $N_f$, and $m_i/v_\Sigma$.
Note that, since $\delta\otheta$ can easily be made smaller than $10^{-2}$ by taking the $a_i^{(\prime)}$ and $c_{1,2}$ couplings in \Eq{eq:delta_theta} to be sufficiently small, this is a conservative constraint.
In \Fig{fig:uv_instanton} we show this largest value of $v_\Sigma$ such that $m_{a,\mathrm{UV}}/m_{a,0} = 10^{-4}$, for different choices of $N_f$ and the ratio $m_i / v_\Sigma$.
Unsurprisingly, small couplings are necessary in order for $v_\Sigma$ to be significantly above the $\TeV$ scale.
Henceforth we take $\alpha' = 0.02$ as a benchmark; the effectiveness of the hidden Witten effect as a function of $\alpha'$ is studied as part of \Sec{sec:mono}.

\subsection{Condition {\it iii}: Monopole disappearance}
\label{subsec:mono_disap}

Let us now turn our attention to the dynamics responsible for the hidden monopoles going away.
For our $\GG' = \Up$ example, the easiest way to do this is with a single charged complex scalar $\phi$.
We assume its potential is such that it develops a non-zero VEV $\vev{\phi} = v_\phi/\sqrt{2}$, thereby Higgsing $\Up$.%
\footnote{Reference \cite{Nomura:2015xil} considered the possibility of leaving $\Up$ unbroken, but having the monopoles be dyons of an additional gauge group $\tilde{\GG}$ to which the axion does not couple.
At the $\tilde{\GG}$ confinement scale $\tilde{\Lambda}$, the $\tilde{\GG}$-color charged monopoles also confine, forming hadronic bound states either among themselves or by capturing $\tilde{\GG}$ quarks.
Additionally, hydrogen-like bound states between the monopoles and electrically-charged $\tilde{\GG}$ quarks can be formed even before $\tilde{\GG}$ confinement \cite{Hook:2024vhf}.
At any rate, once confinement sets in, the resulting final state is a monopole enclosed in a hadron-like object, with a typically-large hadronic annihilation cross $\sigma_\ann \sim 1/\tilde{\Lambda}^2$.
Reference \cite{Nomura:2015xil} suggested that this cross section could be sufficiently large to efficiently annihilate \MMo~pairs.
However, we note that the $r_{\rm had} \sim 1/\tilde{\Lambda}$ size of the hadron and $r_c \sim 1/(\alpha' m_M)$ monopole core are in general vastly different length scales.
Therefore, we expect that annihilations via this cross section will result in the monopoles shedding their ``hadronic'' envelope, while leaving the much smaller monopole cores intact.
If the monopole cores are brought to within the capture radius, \ie, $r_{\rm cap} \sim h^2 / T \sim 1/(e^{\prime 2} T) \gg r_{\rm had}$ (or, equivalently, $T \ll \tilde{\Lambda}/e^{\prime 2}$), the monopole and antimonopole will orbit one another, spiraling ever more closely as they lose energy via radiation emission, and eventually annihilating \cite{Zeldovich:1978wj,Preskill:1979zi,Goldman:1980sn,Vilenkin:2000jqa}.
However, since we expect that $T \sim \tilde{\Lambda}$ when these monopole-hadrons are first formed, this condition requires very large $e'$ couplings.
We leave a detailed analysis of this possibility to future work.}
The $\Up$ SSB gives the $\phi$ boson a mass $m_\phi = \sqrt{\lambda_\phi} v_\phi$, where $\lambda_\phi$ is the $\phi$ quartic coupling.
The hidden photon also acquires a mass $m_{\gamma'} = e' v_\phi$, leading to the screening of hidden electric charges and the confinement of hidden magnetic charges.
Strings made of magnetic flux tubes are then formed between monopole--antimonopole pairs (\MMo) \cite{Nielsen:1973cs,Langacker:1980kd,Lazarides:1981fv,Vilenkin:1982hm}, which shrink in size as they dissipate energy, either by emitting radiation or via interactions with the surrounding thermal bath.
After some time all of the string energy is dissipated, resulting in \MMo~annihilations.

Since for temperatures below $v_\phi$, there are no more light HS particles (barring a model more complex than our benchmark), the energy dissipation of the monopole--antimonopole--string (\MMoS) system must involve interactions with the SM plasma.
Following Ref.~\cite{Nomura:2015xil}, let $\epsilon$ denote the dominant dimensionless coupling between the HS and the SM.
Examples of such couplings include operators such as the Higgs portal $\epsilon \abs{\phi}^2 \abs{H}^2$, or the (hypercharge) photon--hidden photon kinetic mixing $\epsilon F^{\mu\nu} F'_{\mu\nu}$ (which can arise from the irrelevant operator $F^{\mu\nu} \Sigma^i W^{\prime i}_{\mu\nu} / M$) \cite{Brummer:2009cs,Brummer:2009oul}.
Henceforth we consider only the photon kinetic mixing.
One can then find that the main processes through which the \MMoS~system loses energy are: friction with the SM thermal bath at a rate given by \cite{Zeldovich:1978wj,Preskill:1979zi,Goldman:1980sn,Vilenkin:2000jqa}
\beq\label{eq:loss_friction}
    \dot{E}_\fr \sim - \epsilon^2 T^2 v^2 \ ,
\eeq
and, if the strings are ``cuspy'' enough, SM particle production at a rate \cite{Srednicki:1986xg,Vilenkin:2000jqa}
\beq\label{eq:loss_production}
    \dot{E}_\pp \sim - \frac{\epsilon^2 v_\phi^4 \delta^3 L}{\xi_\phi^2} \ .
\eeq
In these expressions $d_M \sim n_M^{-1/3}$ is the typical inter-monopole separation, $v$ is the typical monopole velocity induced by the string tension pulling on them, $\delta$ is the string thickness, $\xi_\phi$ is the $\phi$ correlation length, and $L$ is the string length.
The ``cuspiness'' of the string is determined by whether the inter-monopole distance is larger than the correlation length or not.
If $\xi_\phi > d_M$, then $L = d_M$, whereas in the opposite case the string is made up of a random walk of string segments of length $\xi_\phi$.
Thus, $L \sim \max[ d_M, d_M^2/\xi ]$.
The rates $\dot{E}_\fr$ and $\dot{E}_\pp$ must be compared to the energy stored in the string, namely $E_\str \approx \TT L$, where $\TT \approx \pi v_\phi^2$ is the string tension \cite{Nielsen:1973cs,Forster:1974ga}.%
\footnote{Since cosmic strings are relativistic, their tension is equal to their mass per unit length.}

When $\vert\dot{E}\vert/E_\str$ surpasses the Hubble expansion rate $H$, the energy dissipation becomes efficient; we can use this as an estimate of the time at which the string has shrunk to a point and the \MMo~pairs annihilate.
As first shown in Ref.~\cite{Nomura:2015xil}, these two rates are very fast, unless $\epsilon$ be sufficiently small.
This means that \MMo~annihilations take place soon after string formation, in less than a Hubble time.
If the energy scales of monopole and string formation (roughly speaking $v_\Sigma$ and $v_\phi$, respectively) are distinct enough, there will be enough time for the Witten effect to relax the axion misalignment angle by the $D$ dilution factor required by \Eq{eq:D_value}.
However, if these two scales are not very different, a very small value of $\epsilon$ is necessary in order to delay \MMo~annihilations and give the axion enough time to relax.
As we show in the next subsection, $v_\Sigma$ and $v_\phi$ may indeed be very close, unless our benchmark model be supplemented with an additional mechanism that allows a broad separation of scales.
We postpone a more detailed study of the dynamics of \MMo~annihilation until \Sec{sec:mono}.

\subsection{The Problem with Scalars, and Three Model Versions}
\label{subsec:scalars}

Since $\GG'$ must be part of the bigger $\overline{\GG'}$ in order for monopoles to exist, the $\GG'$-breaking scalar $\phi$ must also be part of a higher-dimensional irrep of $\overline{\GG'}$.
This will generally lead to additional degrees of freedom, as well as additional couplings between $\phi$ and the field $\Sigma$ responsible for Higgsing $\overline{\GG'} \to \GG'$.
In our concrete example, where $\overline{\GG'} = \SUp$, $\Sigma$ can be in an adjoint $\mathbf{3}$ irrep, and $\phi$ in a $\mathbf{2}$ irrep.
Given our definition $e' \equiv g_2' / 2$, the up and down components $\phi_{u,d}$ of $\mathbf{2}$ have $\pm 1$ $\Up$ charges, while the massive $\SUp$ bosons that ``ate'' the $\Sigma$ Goldstones have $\pm 2$ $\Up$ charges.

The most general renormalizable potential for $\Sigma$ and $\phi$ in our benchmark model is therefore
\beq\label{eq:V_scalar}
    \VV(\Sigma, \phi) = - \mu_\Sigma^2 \, \tr\Sigma^2 + \lambda_\Sigma' \, \bl( \tr\Sigma^2 \br)^2 - \mu_\phi^2 \, \abs{\phi}^2 + \lambda_\phi' \, \abs{\phi}^4 + \mu \, \phi^\dagger \Sigma \phi + \tilde{\mu} \, \phi^\dagger \Sigma \tilde{\phi} + \lambda_{\phi\Sigma} \, \abs{\phi}^2 \tr\Sigma^2 + \text{h.c.} \ .
\eeq
where $\tilde{\phi} \equiv i \sigma^2 \phi$; $\phi^\dagger \tilde{\phi}$ vanishes identically. The presence of {\it two} $\Up$-charged scalars in the doublet $\phi$, as well as the mixing terms $\phi^\dagger \Sigma \phi$, $\phi^\dagger \Sigma \tilde{\phi}$, and $\abs{\phi}^2 \tr\Sigma^2$ allowed by gauge invariance, complicate the spectrum of the theory, as well as the dependence of the two SSB scales $v_\Sigma$ and $v_\phi$ on the parameters of the potential.
What is more, these mixing terms will typically lift $v_\phi$ to be close to the scale $v_\Sigma$ of $\SUp \to \Up$ breaking.
This is true even if we were to forbid the $\mu$ and $\tilde{\mu}$ terms in \Eq{eq:V_scalar}, by claiming they are soft-breaking terms of a discrete $\mathbb{Z}_2$ symmetry under which $\Sigma$ transforms non-trivially.
As discussed at the end of the previous subsection, the closeness of the $\Up$ and $\SUp$ SSB scales means that the monopoles are created when the Universe was at a temperature $T_M \sim v_\Sigma$ only to disappear shortly after at temperatures $T_S \sim v_\phi \lesssim v_\Sigma$ when the monopole strings are formed; this is true unless the strings take a very long time to dissipate their energy.
This in turn means that there is not enough time for the monopole-induced axion potential to relax the misalignment angle, thereby defeating the purpose of the axion--monopole coupling in the first place.

In principle, a partial tree-level cancellation between the appropriate terms can result in the necessary $v_\phi \ll v_\Sigma$, at the cost of tuning parameters.
This problem and its unsatisfactory solution are reminiscent of the doublet-triplet splitting problem for the colored Higgs partners in non-supersymmetric GUTs.
To avoid this, we could have one or both scalar VEVs arise from fermion condensation in an additional technicolor sector of a confining $\SU{N}$ gauge group.
The dangerous mixing operators would actually be higher-dimensional, suppressed by the $\SU{N}$ confining scale.
The problem with this is that, if $v_\phi$ arises from just such a condensation, the fermions in question can have a mass only as large as $v_\phi$ itself.
Such light, charged fermions will screen the hidden electric charge of the dyons, increasing $r_a$ in \Eq{eq:beta_axion}, and thereby suppressing their coupling to the axion, once again undoing our work.
Of course, a model where $\phi$ is elementary while $\Sigma$ is composite does not have such a problem.
For example, if $\overline{\GG'} = \SU{3}$, condensation of technifermions in vector-like $(\mathbf{N}, \anti{\mathbf{N}})$ fundamental irreps of $\SU{N}$ but chirally charged in the fundamental $\mathbf{3}$ irrep of $\SU{3}$, can form a scalar $\Sigma$ in the $\mathbf{6}$ symmetric irrep of $\SU{3}$, thereby breaking $\SU{3} \to \SO{3}$ and yielding $\mathbb{Z}_2$-monopoles.
Two elementary scalars $\phi_{1,2}$ in the $\mathbf{3}$ fundamental irrep of the original $\SU{3}$ can further break $\SO{3} \to 1$, leading to monopole confinement and their subsequent annihilation, while mixing terms such as $\tr(\Sigma^2) \abs{\phi}^2$ are higher-dimensional and therefore suppressed.
Alternatively, supersymmetry can also stabilize the potential of the $\Up$-breaking $\phi$.
However, this will also inevitably introduce light fermions in the guise of the $\tilde{\phi}_{L,R}$ higgsinos, who will also be charged under $\Up$ while having a mass $m_{\tilde{\phi}} = m_\phi \sim v_\phi$ \cite{Nomura:2015xil}.

Difficulties in the construction of models where the symmetry breaking pattern $\overline{\GG'} \to \GG' \to 1$ takes place at naturally very different scales are nothing new, and can be thought of as part of the naturalness problem which elementary scalar fields are known to have.
As we have seen, while composite and supersymmetric models can address this issue, they typically introduce light fermions, which end up screening the dyonic hidden electric field ultimately responsible for the relaxation of the axion misalignment angle.

In addition to these problems in the scalar sector responsible for the required breaking patterns of the hidden gauge symmetries, the NB and PQ scalars have their own issues.
The NB scalar $\eta$, apart from suffering from a naturalness problem, will also contribute to the electroweak hierarchy problem through its $c_{1,2}$ couplings to the SM Higgs $H$ \cite{Dine:2015jga}.
Furthermore, the PQ scalar $\Phi$ is known to suffer from the so-called ``quality problem'', where higher-dimensional operators of the form $\abs{\Phi}^{4+n}/\MPl^n$ can spoil the axion solution to the strong CP problem.
Supersymmetry may address some or all of these issues, but it will needlessly complicate the picture beyond the purposes of this paper with the introduction of many new degrees of freedom. 

Another way to improve the quality of the PQ sector without introducing light degrees of freedom is by considering composite axion models~\cite{Kim:1984pt,Choi:1985cb,Kaplan:1985dv}. 
In these models the axion is not a fundamental particle, but rather a composite pion-like object of some gauge group $G$.
Because the axion is a composite, its decay constant is given by the confinement scale of a gauge group $G$, similar to the case of the pion decay constant being related to the QCD condensate~\cite{Witten:1980sp}.
As the axion decay constant is generated by the physics of dimensional transmutation, the composite axion is more protected against higher-order operators spoiling its quality than in the DFSZ or KSVZ models~\cite{Randall:1992ut,Dobrescu:1996jp}. 
For instance, one could consider a confining gauge group $G = \SU{\mathcal N}$ with massless vector-like fermions $\psi$ and $\xi$ transforming under $\SU{\mathcal N}$, $\SUc$ and $\SUp$ as
$\psi = (\Box, 3, R_\psi)$ and $\xi = (\Box, 1, R_\xi)$. 
When $\SU{\mathcal N}$ confines, the global symmetry is spontaneously broken to a smaller group preserving both $\SUc$ color and $\SUp$ color.
The $\SUc$ and $\SUp$ singlet combination is identified with the axion, with its corresponding current being anomalous under both gauge groups.
This combination can be found easily for $R_\psi \neq R_\xi$.
All other non-axion composites get a mass of the order of confinement scale through the perturbative gluon and $\SUp$ gauge boson loops, leaving no light particles in the spectrum.  
Exploring the intricate interplay of the composite axion and the Witten effect requires a detailed study and is beyond the scope of the present work.

Throughout the rest of this paper, we leave the naturalness and axion quality problems of all our scalar fields to future work.
We choose to focus instead on the phenomenology of the basic benchmark model under consideration, namely $\SUp \xrightarrow[]{v_\Sigma} \Up \xrightarrow[]{v_\phi} 1$ broken through a $\SUp$ adjoint $\Sigma$ and a $\SUp$ fundamental $\phi$, with independent VEVs $v_\Sigma$ and $v_\phi$ that are as different as necessary in order for the relaxation of the misalignment angle to reach the required levels.
More concretely, we will study three different versions of our benchmark model:
\begin{itemize}
    \item {\bf baseline}: the $\Sigma$ and $\phi$ potentials are such that $v_\phi \ll v_\Sigma$ (perhaps involving parameter fine-tuning); the lack of additional fermionic degrees of freedom to stabilize the scalar potentials means that $r_a = r_c$,
    \item {\bf stabilized}: $v_\phi \ll v_\Sigma$ but the $\phi$ potential is stabilized through some mechanism (\eg, supersymmetry) requiring the addition of light fermions; this means that $r_a = m_f^{-1} > r_c$), and
    \item {\bf realistic}: here $v_\phi \lesssim v_\Sigma$. In this case the relaxation of the axion misalignment angle takes place with a Higgsed $\Up$, and thus it is necessary that the monopole strings take long to dissipate their energy.
\end{itemize}
Finally, a comment on notation.
From now on we use $\lambda_\Sigma$ and $\lambda_\phi$ to denote ``effective'' quartic couplings of the $\Sigma$ and $\phi$ scalars, even though in general we cannot neatly identify them with $\lambda_\Sigma'$ and $\lambda_\phi'$ in \Eq{eq:V_scalar} due to the mixing terms.
This shorthand will be useful below when we discuss in more detail the critical temperatures and correlation lengths associated with the SSB of $\SUp$ and $\Up$, which in general depend on such Higgs self-interactions.
In fact, we can simply {\it define} $\lambda_\phi \equiv m_\phi^2/v_\phi^2$, where $m_\phi$ is the mass of the lightest boson in $\phi$.

\section{Monopole--axion Dynamics}
\label{sec:mono}

The total energy density stored in the axion--monopole interactions, coming from a population of monopoles and antimonopoles with number density $n_M = n_{M+} + n_{M-}$, can be obtained from \Eq{eq:axion_dyon_potential}:
\beq\label{eq:U_axion}
    U = V_0 n_M = \frac{1}{2} \bl( \frac{2 \beta}{f_a} n_M \br) a^2 \equiv \frac{1}{2} m_{a,M}^2 a^2 \ ,
\eeq
where, based on the discussion found in \Subsec{subsec:nelson_barr}, we have plugged $\othetap = 0$.
It is clear that $m_{a,M}^2 \equiv 2 \beta n_M / f_a$ is the axion mass from the $\VV_M$ monopole contribution, and that it will be playing the role of $m_{a,X}^2$ from \Subsec{subsec:two_mis}.
Let us devote our attention first to the creation of a monopole abundance, its interactions with the axion field via \Eq{eq:U_axion}, and its subsequent disappearance upon $\Up$ Higgsing.

\subsection{Formation and Abundance of Hidden Monopoles}

Monopole formation can take place through different mechanisms.
The one most commonly studied in the literature is the Kibble-Zurek mechanism \cite{Kibble:1976sj,Zurek:1985qw}, in which a phase transition (PT) takes place at a temperature $T_M$.
The exact definition of $T_M$ depends on the nature of the PT in question.
For example, for second-order phase transitions (SOPTs), $T_M$ is commonly taken to be the Ginzburg temperature, defined as that temperature below which thermal fluctuations are unable to undo the topologically non-trivial field configurations that undergird the existence of the monopoles.
For small $\Sigma$ self-couplings $\lambda_\Sigma$, the temperatue $T_M$ is not too different from the critical temperature of the PT, which is $\sim v_\Sigma$ if $\lambda_\Sigma \gtrsim g_2'$, and $\sim \sqrt{\lambda_\Sigma / g_2'^2} \, v_\Sigma$ if $\lambda_\Sigma < g_2'$ \cite{Kibble:1980mv,Vilenkin:2000jqa,Kibble:1980mv}.
For our current purposes we take $T_M$ to be $\OO(1)$ close to the PT critical temperature.
The number density of monopoles at $T_M$ is determined by the correlation length $\xi_M \equiv \xi_\Sigma(T_M)$ of the $\Sigma$ Higgs at the time of the phase transition \cite{Guth:1979bh,Einhorn:1980ym,Kibble:1980mv}:
\beq\label{eq:nM_TM}
    n_M(T_M) \equiv p_0 \, \xi_M^{-3} \ .
\eeq
This means that there are $\kappa_M \equiv p_0/(H(T_M) \, \xi_M)^3$ monopoles per Hubble volume at the time of formation.
The numerical factor $p_0 \sim \OO(0.1)$ is model-dependent, and it corresponds to the probability that the orientation of the $\Sigma$ Higgs is topologically non-trivial at the intersection of its uncorrelated domains, and is thus capable of producing monopoles \cite{Kibble:1976sj}.
Causality demands that $\xi_M < H(T_M)^{-1}$, and thus $\kappa_M \gtrsim 0.1$; it can often be much larger.
For a SOPT, for example, $\xi(T_M)$ is very roughly $\sim 1/T_M$, and thus $\kappa_M \sim p_0 (\mPl / \sqrt{g_{*M}}T_M)^3 \sim 10^{39} \, (10~\TeV / T_M)^3$.
Yet another example is $\kappa_M = (\beta_\PT / v_w)^3$ for a first-order PT (FOPT) with bubble nucleation action $S$, nucleation rate $\beta_\PT \equiv - H \dd S/\dd t$, and bubble wall velocity $v_w$.
A natural benchmark for such a FOPT is $\beta_\PT / v_w \approx 100$ and thus $\kappa_M \approx 10^6$ \cite{Sato:2018nqy,Yang:2022quy}.

Shortly after being formed, attractive interactions between monopoles and antimonopoles will bring them together.
If their interactions with the thermal plasma are efficient enough, the \MMo~pairs lose energy and capture one another in a process known as ``diffusive capture'' \cite{Zeldovich:1978wj,Preskill:1979zi,Goldman:1980sn,Vilenkin:2000jqa}.
Eventually the \MMo~pairs are brought into close proximity and annihilate.
Diffusive capture can only occur if the monopole mean free path $\lambda_\mfp \sim (B T)^{-1} \sqrt{m_M/T}$ is smaller than the monopole capture radius $r_{\rm cap} \sim h^2/T$.
Here $h = 1/g_2' = 1/(2 e')$ is the magnetic charge of the ('t Hooft-Polyakov) monopole \cite{tHooft:1974kcl,Polyakov:1974ek}, and $B$ is a numerical coefficient accounting for the drag force that the plasma particles impart on the monopoles, and is given by \cite{Preskill:1979zi,Goldman:1980sn,Vilenkin:2000jqa}
\beq
    B \equiv B_{\rm HS} + B_{\rm SM} \approx \frac{\pi}{36} C' \sum\limits_{i'} g_{i'} c_{i'} q_{i'}^2 (2 e' h)^2 + B_{\rm SM} \ .
\eeq
The sum runs over the HS relativistic species $i'$ with $\Up$ charge $q_{i'}$.
The $c_{i'}$ prefactor is equal to $1$ ($1/2$) for bosons (fermions), while $g_{i'}$ counts spin degrees of freedom of the $i'$-th species.
The $B_{\rm SM}$ term corresponds to the contribution to this drag force coefficient coming from any possible interactions between the SM particles in the plasma and the hidden monopoles.
For the kinetic mixing $\epsilon F^{\mu\nu} F'_{\mu\nu}$ between $\Ua_Y$ hypercharge and $\Up$ gives $B_{\rm SM} = \epsilon^2 (\alpha_Y/\alpha') (\pi/36) C \sum_{i} g_i c_i q_i^2 (2 e' h)^2$, for particles of hypercharge $q_i$.%
\footnote{
There is a subtlety regarding the plasma friction on the hidden monopoles arising from the $\epsilon F^{\mu\nu}F'_{\mu\nu}$ term.
Such a kinetic mixing results in not only the SM charged particles acquiring a hidden millicharge, but also in the hidden monopoles obtaining a SM millimagnetic charge, of the opposite sign.
The result is that the net Lorentz force of the HS/milli-SM monopole on the SM/milli-HS charged particles is zero.
This can be gleaned from angular momentum considerations: the total angular momentum of the monopole-charged particle system, which is proportional to the product of their couplings, must be quantized; since $\epsilon$ is a continuous parameter, the only possibility is that the net coupling vanishes.
However, once $\Up$ is Higgsed and the magnetic field lines are confined into strings, this argument ceases to be true at distances larger than the thickness of the string, namely the Compton wavelength of the massive hidden photon.
In this case $B_{\rm SM}$ may still play a role.
For the Higgs portal $\epsilon \abs{\phi}^2 \abs{H}^2$ there is no such subtlety: there is no tree-level coupling between the hidden monopole and the SM plasma, and so $B_{\rm SM} \approx 0$.}
The numerical coefficient $C^{(\prime)}$ regulates the monopole-plasma momentum-transfer cross section, and it is roughly given by $C^{(\prime)} \sim \ln(T/(n^{(\prime)} \sigma_{\rm plasma}^{(\prime)}))$, where $n^{(\prime)} \approx \zeta(3) g_*^{(\prime)} T^3 / \pi^2$ and $\sigma_{\rm plasma}^{(\prime)} \sim \alpha^{(\prime)\, 2} T^{-2}$ are the SM (HS) plasma number density and self-scattering cross section respectively.
In our benchmark model, for example, the four degrees of freedom associated with the $\Up$-charged, light $\phi$ particles in the $\mathbf{2}$ irrep of $\SUp$, give $B_{\rm HS} \approx 1.85$ for $\alpha' = 0.1$, and $B_{\rm HS} \approx 3.47$ for $\alpha' = 0.01$.
For the $\Ua$ kinetic mixing coupling we consider in this paper, we have $B_{\rm SM} \approx 0.1 \, \epsilon^2 / \alpha'$.
Thus, $\lambda_\mfp < r_{\rm cap}$ only when the Universe was hotter than the diffusion temperature
\beq\label{eq:T_diff}
    T_d \sim \frac{m_M}{h^4 B^2} \sim \bl( \frac{\alpha'}{0.02} \br)^{3/4} \bl( \frac{3}{B} \br)^2 \bl( \frac{1}{2 e' h} \br)^4 \bl( \frac{v_\Sigma}{T_M} \br) T_M\ .
\eeq
Thus, since $v_\Sigma$ can be much larger than $T_M$ ($\sim v_\Sigma \, \min[\sqrt{\lambda_\Sigma}/e', 1]$ for a SOPT), $T_d$ is also larger than $T_M$ for $\alpha'$ above a few percent.
Compare with the very different case of GUT monopoles, for which $T_d$ can be four or five orders of magnitude below the GUT scale \cite{Preskill:1979zi,Goldman:1980sn,Vilenkin:2000jqa}.

The above discussion shows that for the parameter space of interest, \MMo~annihilations do not reduce the initial monopole abundance given by \Eq{eq:nM_TM}.
Therefore, we can parametrize the monopole abundance as
\bea\label{eq:nM}
    \frac{n_M}{s} & = & \frac{p_0}{\xi_M^3 \, s_M} = \kappa_M \, \frac{H(T_M)^3}{s_M} \ ,
\eea
where $s = 2 \pi^2 g_{*s} T^3 / 45$ is the entropy of the Universe, with $g_{*s}$ counting the number of relativistic degrees of freedom in entropy.
If the comoving entropy in the SM and HS are conserved, then $n_M/s$ is constant.
At sufficiently high temperatures, the particles in the standard and hidden sectors are relativistic, and thus we have $g_{*s} = 119.75$, which includes $106.75$ from the SM, and $2 \cdot 3 \, (W^{\prime i}) + 3 \, (\Sigma^i) + 2 \cdot 2 \, (\phi) = 13$ degrees of freedom from the HS.%
\footnote{Technically the $W^{\prime \pm}$ and $\Sigma^0$ particles become non-relativistic around $T_M \sim v_\Sigma$; therefore $g_{*sM} \equiv g_{*s}(T_M)$ is somewhat smaller than 119.75.}
In any case, \Eqs{eq:U_axion}{eq:nM} show that $m_M^2 \sim T^3$ and, therefore, with $q = 3$, \Eq{eq:D_def} gives $D \sim T^{3/4}$.

\subsection{Axion-Monopole Couplings}

In order for the axion relaxation to take place, it is necessary that the condition for the onset of axion oscillations, namely $m_{a,M} > 3 H$, be satisfied at some point.
Since $n_M \sim T^3$ the monopoles behave as a matter-like energy density component: given sufficient time they will lead to an era of early matter domination, with the Hubble rate scaling like $H \propto \sqrt{\rho_M} = \sqrt{m_M n_M} \sim T^{3/2}$.
Since $m_{a,M} \sim T^{3/2}$ as well, the ratio $m_{a,M} / H$ is constant during matter domination, and thus the axions {\it must} start to oscillate before this era, or they never will.
Therefore, a necessary but not sufficient condition for the onset of axion oscillations is $m_{a,M} > 3 H(\rho_M) = 3 \sqrt{\rho_M / (3 \mPl^2)}$, or equivalently $2 \beta / f_a > 3 m_M/\mPl^2$.
In terms of the axion decay constant, this consistency condition reads
\beq\label{eq:fa_MD}
    f_a < f_a^{\rm max} \equiv \frac{\alpha'^{3/4} \, E'/N}{4 \sqrt{3} \pi^{5/4}} \frac{\mPl}{\sqrt{r_a \, v_\Sigma}} \approx 2.7 \times 10^{16}~\GeV \, \bl( \frac{\alpha'}{0.02} \br) \bl( \frac{E'/N}{6} \br) \sqrt{\frac{r_c}{r_a}} \ ,
\eeq
where we have used $r_c v_\Sigma \approx 1/(4 \sqrt{\pi \alpha'})$.
If $r_a = r_c$, as in the baseline version of our benchmark model, this consistency condition is trivially satisfied.
However, in the stabilized benchmark model the presence of light charged fermions makes $r_c / r_a < 1$, thereby tightening the \Eq{eq:fa_MD} constraint.
Note that the presence of additional matter-like relics (\eg, the massive $W^{\prime \pm}$ and $\Sigma$ particles) will only ensure an even earlier onset of matter domination, tightening the consistency condition on $f_a$.
However, these relics promptly decay into the lighter hidden particles, such as the hidden photon $A'$ and the $\SUp$ doublet $\phi$.
If these particles have significant portal couplings to the SM (the $\epsilon$ coupling of \Subsec{subsec:mono_disap}, necessary for energy dissipation by the \MMo~string), they will in turn decay into SM states.
This is the case of the baseline and stabilized versions of our benchmark models.
On the other hand, in the realistic model $v_\phi \lesssim v_\Sigma$ and the $\epsilon$ couplings of the HS to the SM must of necessity be very small, in order to give enough time for the relaxation of the axion misalignment angle.
In this case there will always be a lightest massive particle from the HS which will contribute to the early onset of matter domination, and against which the axion mass $m_{a,M}$ must compete.

As long as the $f_a$ consistency condition is satisfied, the monopole-induced axion oscillations will begin during the radiation-dominated era, at a temperature that can be determined by requiring $m_{a,M}(T_a) = 3 H(T_a)$, and is thus given by
\bea
    T_a & \approx & \frac{5}{8 \pi^4} \frac{p_0 (E'/N)^2 \, \alpha'}{g_{*M}} \frac{\mPl^2}{f_a^2 \, r_a \, \bl( \xi_M T_M \br)^3} \nonumber \\
    & \approx & 10~\TeV \, \bl( \frac{p_0}{0.1} \br) \bl( \frac{119.75}{g_{*M}} \br) \bl( \frac{\alpha'}{0.02} \br) \bl( \frac{E'/N}{6} \br)^2 \bl( \frac{4.8 \times 10^{15}~\GeV}{f_a} \br)^2 \bl( \frac{1}{\xi_M T_M} \br)^3 \bl( \frac{1}{r_a T_M} \br) \bl( \frac{T_M}{10~\TeV} \br) \ . \label{eq:T_oscu}
\eea
Note that if $r_a = r_c$, then $r_a T_M \approx \sqrt{0.02/\alpha'} \, (T_M/v_\Sigma)$.
If $T_a$ were to be larger than $T_M$ then the axion field begins to oscillate as soon as the monopoles are created; in other words $T_\oscu = \min[T_a, T_M]$.

\subsection{String Formation and Monopole Annihilation}

As the Universe continues cooling down, the $\phi$ scalar will develop a VEV and Higgs $\Up$ at some critical temperature.
This spontaneous symmetry breaking generates strings of magnetic flux connecting monopoles with antimonopoles.
At some temperature $T_S$ thermal fluctuations are unable to undo the topologically non-trivial winding of the scalar field $\phi$, and the string formation ends.
As in the case of $T_M$, the precise value of $T_S$ depends on the details of the PT responsible for breaking $\Up$.
For example, for a single charged scalar $\phi$ and a single gauge boson $\Up$, and couplings not too big, a SOPT yields $T_S$ close to the critical temperature, namely $T_S \approx \sqrt{3 \lambda_\phi / (\lambda_\phi + 3e'^2/4)} \, v_\phi$ \cite{Nielsen:1973cs,Vilenkin:2000jqa}.
In our benchmark model, however, the fact that $\phi$ is part of an $\SUp$ doublet, as well as the existence of the $W'^\pm$ and $\Sigma^0$ bosons, complicates the picture.
Nevertheless, one still expects that for an SOPT and small couplings $T_S \sim v_\phi \min[\sqrt{\lambda_\phi}/e', 1]$.

Once the strings are formed, they begin to lose energy at the rates given by \Eqs{eq:loss_friction}{eq:loss_production}.
As discussed in the previous section, and shown in Ref.~\cite{Nomura:2015xil}, this can happen in less than one Hubble time as long as the HS coupling to the SM, $\epsilon$, is not too small.
The process that dominates the string energy loss depends on whether the $\phi$ correlation length at the time of formation, $\xi_S \equiv \xi_\phi(T_S)$, is shorter or longer than the average inter-monopole distance, $d_M$:
\beq\label{eq:dM_xiS}
    \frac{d_M}{\xi_S} \approx \bl( \frac{g_{*M}}{g_{*S} p_0} \br)^{1/3} \frac{\xi_M T_M}{\xi_S T_S} \sim \frac{\xi_M T_M}{\xi_S T_S} \ .
\eeq
Both $d_M/\xi_S < 1$ and $d_M / \xi_S > 1$ can be achieved with a modest hierarchy between the $\lambda_\phi$ and $\lambda_\Sigma$ couplings since they, along with $e'$, determine the parameter combinations $\xi_S T_S$ and $\xi_M T_M$ in a SOPT, respectively.

If $d_M / \xi_S < 1$, the strings connecting the \MMo~pairs are straight, with an average length $L = d_M$, making plasma friction the dominant energy loss process.
All the HS particles become non-relativistic at around $T_S$; as they cool, they slow down and cease to exert significant friction on the \MMo~pair.
Therefore, only the relativistic SM particles contribute to the energy loss.
The rate associated with this process at $T_S$ is
\beq
    \Gamma_\fr \equiv \frac{\vert\dot{E}_\fr\vert}{E_\str} \approx \frac{B_{\rm SM} T_S^2 v^2}{\TT d_M} \gtrsim 0.06 \, \frac{\epsilon^2 T_S^2}{\sqrt{\alpha'} v_\Sigma} \ ,
\eeq
where in the last expression we have used $v \sim \min[\sqrt{\TT d_M / m_M}, 1]$ as the typical monopole velocity caused by the string tension $\TT$ pulling on it.
Since $\sqrt{\TT d_M / m_M} \sim \alpha'^{1/4} \sqrt{v_\phi / v_\Sigma} \sqrt{T_M \xi_M v_\phi / T_S}$ at temperature $T_S$, even a moderate $v_\phi / v_\Sigma$ hierarchy, with $T_S \sim v_\phi$, and $\xi_M^{-1} \sim T_M \sim v_\Sigma$, yields $v < 1$.
Therefore, as long as $\epsilon \gtrsim (g_{*S} \alpha')^{1/4} \sqrt{v_\Sigma / \mPl} \sim 10^{-7} \, (\alpha'/0.02)^{1/4} \sqrt{v_\Sigma / 10~\TeV}$ the energy loss process will take less than a Hubble time.
If this is the case then the strings will quickly dissipate all of their energy and the \MMo~pairs will annihilate, leading to $T_\finu \approx T_S$.
On the other hand, if $\epsilon$ is much smaller than this threshold, then the energy dissipation of the \MMoS~system may take several Hubble times, and thus $T_\finu \ll T_S$.

Alternatively, if $d_M / \xi_S > 1$ then the \MMo~strings are formed in a Brownian way.
This results in very cuspy strings, and string particle production becomes important.
The associated rate is
\beq
    \Gamma_\pp \equiv \frac{\vert\dot{E}_\pp\vert}{E_\str} \approx \frac{\epsilon^2 v_\phi^4 \delta^3}{\TT \xi_\phi^2} \ .
\eeq
The string thickness $\delta$ cutting off the energy of the emitted particles depends on the nature of the SM--hidden interaction.
For the photon mixing coupling $\epsilon F'_{\mu\nu} F^{\mu\nu}$ we take $\delta \approx m_{\gamma'}^{-1}$, the thickness of the string's inner magnetic flux tube.
For temperatures below $T_S$, the temperature corrections to these quantities become negligible, and therefore we have roughly $\Gamma_\pp \sim \epsilon^2 v_\phi^2 m_\phi^2 \delta^3 \sim \epsilon^2 v_\phi$.
This is much faster than the radiation-era Hubble rate at $T_S$, $H \sim T_S^2 / \mPl \sim v_\phi^2 / \mPl$, as long as $\epsilon \gtrsim \sqrt{v_\phi / \mPl}$, barring extremely large hierarchies between the $\lambda_\phi$ and $e'^2$ couplings.
As we show in the next section, $v_\phi \sim 100~\GeV$ is a reasonable benchmark that allows for large $f_a$ axion DM; the above condition becomes $\epsilon \gtrsim 10^{-8}$.
Since $\Gamma_\pp > H(T_S)$ for values of $\epsilon$ above this threshold, we find that the cuspy string will quickly dissipate its energy, before one Hubble time.
Alternatively, smaller $\epsilon$ values lead to the energy of the \MMoS~system taking several Hubble times to dissipate.

\section{Results}
\label{sec:results}

In this section we turn our attention to the careful study of the parameter space of each of the versions of our benchmark model in turn.
Given $T_\oscu = \min[T_a, T_M]$ (see \Eq{eq:T_oscu}), and $T_\finu$ (which depends on whether the \MMoS~system loses energy quickly or not), we can compute the dilution factor $D$ of \Eq{eq:D_def}, and find the axion DM values of $f_a$ that our dynamical misalignment mechanism can achieve.

\subsection{Baseline model case}

As first mentioned in the previous section, the baseline model has $v_\phi \ll v_\Sigma$ and $r_a = r_c \approx 1/(2 e' v_\Sigma)$.
With this in mind, \Eq{eq:T_oscu} turns into
\beq
    \frac{T_a}{10~\TeV} \approx \bl( \frac{p_0}{0.1} \br) \bl( \frac{119.75}{g_{*M}} \br) \bl( \frac{\alpha'}{0.02} \br)^{3/2} \bl( \frac{E'/N}{6} \br)^2 \bl( \frac{4.8 \times 10^{15}~\GeV}{f_a} \br)^2 \bl( \frac{1}{\xi_M T_M} \br)^3 \bl( \frac{v_\Sigma}{T_M} \br) \bl( \frac{T_M}{10~\TeV} \br) \ . \label{eq:baseline_T_oscu}
\eeq

In this case we assume $\epsilon \gtrsim 10^{-7} \, (\alpha' / 0.02)^{1/4} \sqrt{v_\Sigma / 10~\TeV}$.
As discussed above, this coupling is large enough such that the \MMoS~system completely disappears shortly after its formation, ensuring $T_\finu \approx T_S$.
Based on this and \Eq{eq:baseline_T_oscu}, \Eqs{eq:D_def}{eq:D_value} can be leveraged to find the values of $f_a$ for which the monopole-induced oscillations guarantee the axion abundance will match that of dark matter:
\bea\label{eq:fa_baseline_1}
    \text{for }T_a < T_M : \ \frac{f_a}{4.8 \times 10^{15}~\GeV} & \approx & \bl( \sqrt{\frac{\Omega_a h^2}{0.12}} \frac{2.15}{\theta_\infl} \br)^{0.48} \bl( \frac{p_0}{0.1} \frac{119.75}{g_{*M}} \br)^{0.36} \bl( \frac{\alpha'}{0.02} \br)^{0.54} \bl( \frac{E'/N}{6} \br)^{0.72} \nonumber\\
    && \times \bl( \frac{1}{T_M \xi_M} \br)^{1.1} \bl( \frac{4.2 \times 10^{-4}}{T_S/v_\Sigma} \br)^{0.36} \ , \\
    \text{for } T_a > T_M: \ \frac{f_a}{4.8 \times 10^{15}~\GeV} & \approx & \bl( \sqrt{\frac{\Omega_a h^2}{0.12}} \frac{2.15}{\theta_\infl} \br)^{1.7} \bl( \frac{4.2 \times 10^{-4}}{T_S/T_M} \br)^{1.3} \ ,\label{eq:fa_baseline_2}
\eea
where we have used the benchmark $r_c T_S \approx 4.1 \times 10^{-4} \, \sqrt{0.02/\alpha'} ((T_S/v_\Sigma) / 4.2 \times 10^{-4})$.
For a SOPT, we have roughly $T_M \sim v_\Sigma$ and $T_S \sim v_\phi$, and thus we see that we need at least a hierarchy of $v_\phi / v_\Sigma \sim 10^{-4}$ for an axion with $f_a \approx 5 \times 10^{15}~\GeV$ to be the entirety of DM.
For our $v_\Sigma \sim 10~\TeV$ benchmark this means $v_\phi \sim 1~\GeV$.

\begin{figure}[t!]
\centering
\includegraphics[width=0.49\linewidth]{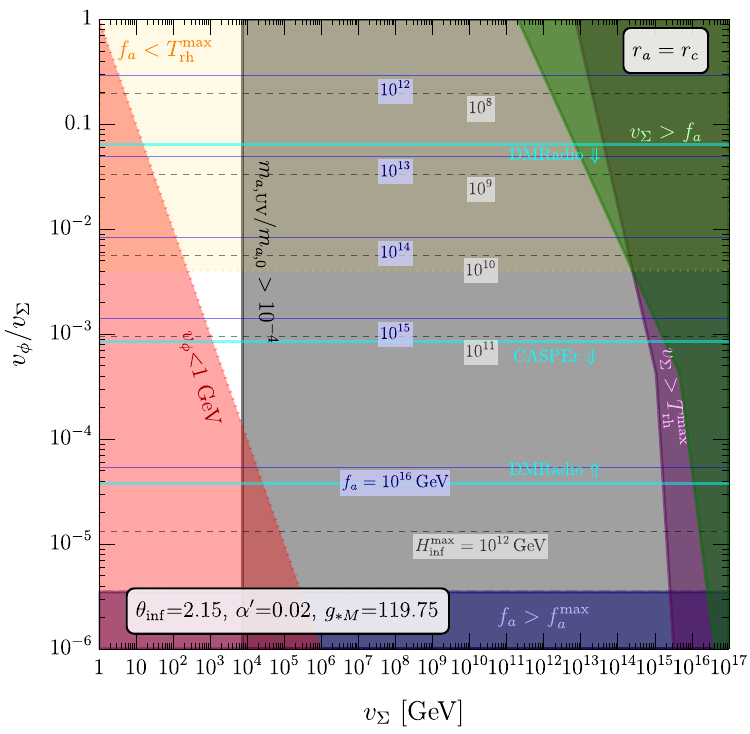}
\includegraphics[width=0.49\linewidth]{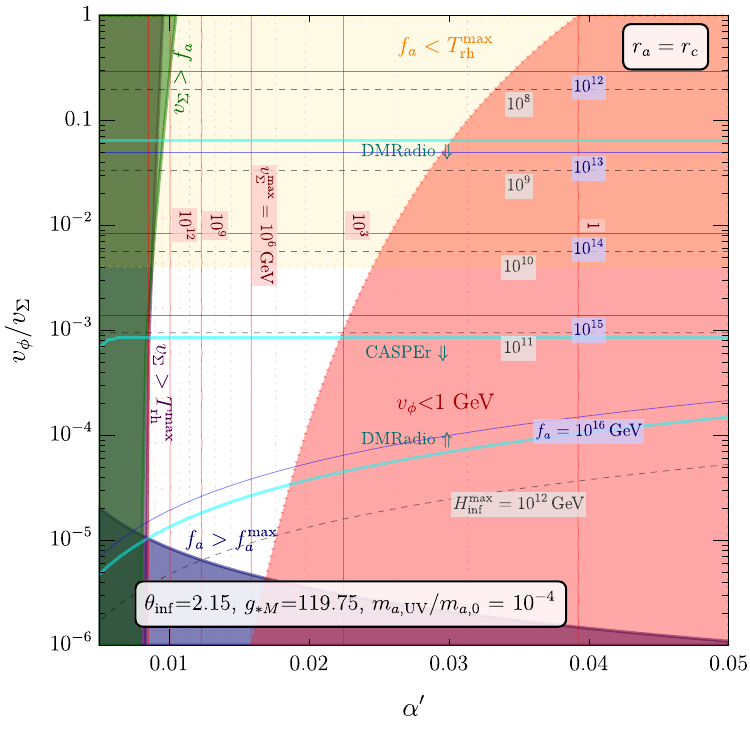}
\caption{The parameter space of the baseline model ($r_a = r_c$, $\epsilon \gtrsim 10^{-7} \, (\alpha' / 0.02)^{1/4} \sqrt{v_\Sigma / 10~\TeV}$).
We focus on $v_\Sigma$ vs. $v_\phi/v_\Sigma$ with $\alpha' = 0.02$ in left panel, and on $\alpha'$ vs. $v_\phi / v_\Sigma$ in the right panel, with $v_\Sigma$ taken from \Fig{fig:uv_instanton} with $N_f = 0$.
We assume $\xi_M^{-1} \approx T_M \approx v_\Sigma$ and $T_S \approx v_\phi^{-1}$, and take $p_0 = 0.1$, $E'/N = 6$, $\theta_\infl = 2.15$, and $g_{*M} = 119.75$.
The blue contours denote the axion DM values of $f_a$ allowed by the model (\Eqs{eq:fa_baseline_1}{eq:fa_baseline_1}), with the associated maximum values of $\Hinf^{\rm max}$ consistent with isocurvature constraints (\Eq{eq:iso_Hinfl}) depicted as gray dashed lines.
In blue is the region for which the hidden Witten effect cannot make the axion field oscillate (\Eq{eq:fa_MD}).
Shaded in red are those points for which $v_\phi < 1~\GeV$ and thus entropy dumps from the HS may upset the standard QCD misalignment mechanism.
In purple and green are those points for which a pre-inflationary axion and a post-reheating formation of monopoles cannot be achieved.
In yellow is that region for which the reheating temperature must be significantly smaller than its maximum value $T_\rh^{\rm max} \approx E_{\rm infl}^{\rm max}$ of instantaneous reheating, in order to avoid restoring the PQ-symmetry.
The cyan contours show the sensitivity reach of DMRadio \cite{DMRadio:2022pkf} and CASPEr-electric \cite{Budker:2013hfa,JacksonKimball:2017elr}.
In the left panel, the gray shaded area has $v_\Sigma$ so large that $m_{a, \mathrm{UV}} > 10^{-4} \, m_{a,0}$, spoiling the axion solution to the strong CP problem if $\delta \otheta \gtrsim 10^{-2}$ in \Eq{eq:delta_theta}.
In the right panel, the red lines denote the values of $v_\Sigma$ such that this bound is saturated.
}
\label{fig:baseline}
\end{figure}

In \Fig{fig:baseline} we depict the parameter space of the baseline model, for which $r_a = r_c$.
We assume $\xi_M^{-1} \approx T_M \approx v_\Sigma$ and $T_S \approx v_\phi^{-1}$, and take $p_0 = 0.1$, $E'/N = 6$, and $\theta_\infl = 2.15$.
We further assume $\epsilon$ is large enough so that $T_\finu \approx T_S$.
In the left panel we show the $v_\Sigma$--$v_\phi/v_\Sigma$ parameter space with fixed $\alpha' = 0.02$, while in the right panel we vary $\alpha'$, at each point taking $v_\Sigma$ to be as large as possible while keeping the contributions to the axion mass coming from UV instantons to be less than $10^{-4}$ that of QCD, as discussed in \Subsec{subsec:axion_mono_couplings} and shown in \Fig{fig:uv_instanton}.%
\footnote{In this and all sections, we ignore the renormalization group running of the $\Up$ coupling $\alpha'$.
This is justified given the small values of $\alpha'$ and the modest number of HS charged particles considered in our paper.
For example, a quick estimate shows that taking $\alpha'(v_\Sigma) = 0.02$ and a single charged hidden fermion results in $\alpha'(10^{-6} \, v_\Sigma) \approx 0.0194$: a change of less than $1\%$ over six orders of magnitude in energy.}
We assume there is no extra vector-like $\SUp$ fermion doublets (\ie, $N_f = 0$), and therefore take $g_{*M} = 119.75$.
In both plots, the blue contours correspond to the values of $f_a$ of \Eqs{eq:fa_baseline_1}{eq:fa_baseline_2}, for which the dilution factor $D$ is such that the axion abundance accounts for 100\% of dark matter (\ie, $\Omega_a h^2 \approx 0.12$).
The largest possible values of the inflationary scale $\Hinf^{\rm max}$, for which the isocurvature bound is satisfied (\Eq{eq:iso_Hinfl}), are shown as dashed gray contours.
The parameter region violating the consistency condition of \Eq{eq:fa_MD} is shown in blue.
In red we show the region for which $v_\phi < 1~\GeV$; these values imply that the non-relativistic HS relics, as well as the \MMoS~system, dump their entropy dangerously close to the QCD phase transition, possibly upsetting the standard QCD axion misalignment mechanism.
In green we show the region for which $v_\Sigma > f_a$; since we assume a pre-inflationary axion and a post-inflationary, post-reheating production of hidden monopoles, such a region is inconsistent.
The purple region is inconsistent with the assumption that the hidden monopoles are produced after reheating, since for points within this area $v_\Sigma$ is larger than the largest possible reheating temperature $T_\rh^{\rm max} \approx E_\infl^{\rm max} = (3 \mPl^2 H_\infl^{\rm max \, 2})^{1/4} \approx 7 \times 10^{-3} \sqrt{\mPl f_a \theta_\infl} \approx 1.1 \times 10^{15} \sqrt{(f_a / (4.8 \times 10^{15}~\GeV)) (\theta_\infl / 2.15)}$, corresponding to the case of instantaneous reheating after inflation.
In the yellow region $f_a < T_\rh^{\rm max}$, that is to say $f_a < 2.5 \times 10^{14}~\GeV \, (\theta_\infl / 2.15)$, which would imply the restoration of PQ-symmetry after reheating.
This is not a strict bound, since it can be relaxed simply by choosing reheating to take place much later than the end of inflation.
In the left panel, where $\alpha' = 0.02$, those values of $v_\Sigma$ for which $m_{a, \mathrm{UV}} > 10^{-4} \, m_{a,0}$ are shaded in gray.
For these points the axion solution to the strong CP problem is spoiled if $\delta \otheta \gtrsim 10^{-2}$ in \Eq{eq:delta_theta}.
In the right panel the condition $m_{a, \mathrm{UV}} = 10^{-4} \, m_{a,0}$ is satisfied everywhere, since we take $v_\Sigma$ to be as large as possible while maintaining this equality.
The corresponding values of $v_\Sigma$ are plotted as the red contours.
In both panels we show, in cyan, the experimental reach of DMRadio ($1.1 \times 10^{16}~\GeV \gtrsim f_a \gtrsim 7.1 \times 10^{12}~\GeV$) \cite{DMRadio:2022pkf} and CASPEr-electric ($f_a \gtrsim 1.9 \times 10^{15}~\GeV$) \cite{Budker:2013hfa,JacksonKimball:2017elr}.

We can see that in the baseline model, the dynamical misalignment mechanism generated by a hidden monopole Witten effect allows axion DM with $f_a$ values larger than $\sim 10^{16}~\GeV$, all while satisfying the axion isocurvature bounds and other consistency checks.
This represents an increase of more than four orders of magnitude compared to what the standard QCD misalignment mechanism predicts.
In particular, the baseline model allows $f_a$ to take any value in DMRadio's sensitivity range, as well as in the lower portion of CASPEr-electric's.
However, the corresponding values of $\Hinf^{\rm max}$ and the inflationary $r$-parameter are beyond the sensitivity of next-generation CMB experiments.
Indeed, the CMB-HD experiment's $1\sigma$ reach for the tensor-to-scalar ratio $r$ ($\sigma(r) = 5 \times 10^{-4}$) \cite{Sehgal:2019ewc,Sehgal:2020yja,CMB-HD:2022bsz} corresponds to $\Hinf^{\rm max} \approx 5.5 \times 10^{12}~\GeV$, and thus $f_a \approx 2 \times 10^{17}~\GeV$, which is well within the region where the axion oscillation consistency condition is violated (blue region of \Fig{fig:baseline}); LiteBIRD will have a similar reach of $\sigma(r) \sim 10^{-3}$ \cite{LiteBIRD:2023zmo}.
However, future experiments may eventually probe the $r$-parameter values predicted by this model.
Note that the scale $v_\Sigma$ of $\SUp$-breaking and monopole generation can span many orders of magnitude, from $\sim 1~\GeV$ to $\sim 10^{15}~\GeV$; however, the hierarchy between this scale and that of $\Up$-breaking and string formation must not be larger than $10^5$.
For $\alpha' \approx 0.02$, the largest $f_a$ can be obtained with $v_\Sigma \sim 10^5~\GeV$ and $v_\phi \sim 1~\GeV$.
For these scales and couplings, the hidden photon has a mass of $m_{\gamma'} \approx 0.5~\GeV$; collider and fixed-target bounds on the photon kinetic mixing ($\epsilon \lesssim \text{few} \times 10^{-4}$) are not stringent enough to rule out this model \cite{Fabbrichesi:2020wbt}.

We reiterate that the baseline model is rather simplistic, since it ignores the question of the radiative stability of the $\phi$ potential.
We take this issue into account in the next subsection, where we consider the stabilized version of our model.

\subsection{Stabilized model case}

Once again we assume that string energy dissipation is efficient enough so that $T_\finu \approx T_S$.
In the stabilized version of our model, the $\phi$ potential is stabilized against corrections from $\Sigma^0$ and $W'^\pm$ contributions.
Thus, $v_\phi \ll v_\Sigma$ (and therefore $T_S \ll T_M$) can be obtained without fine-tuning, but typically at the cost of additional light fermionic degrees of freedom.
For concreteness we consider the supersymmetric case, where SUSY keeps $v_\phi$ smaller than $v_\Sigma$, but at the cost of introducing the $\tilde{\phi}_{L,R}$ hidden higgsinos, with masses $m_{\tilde{\phi}} = m_\phi = \sqrt{\lambda_\phi} \, v_\phi$.
Because of this, $r_a \approx m_\phi^{-1}$, and therefore $r_a T_S \approx T_S / m_\phi$.
Barring a supercooled FOPT for which $T_S$ could be well below the PT critical temperature, we therefore expect $r_a T_S \sim \min[ 1/e', 1/\sqrt{\lambda_\phi}]$.
This scenario was first sketched in Ref.~\cite{Nomura:2015xil}, although working under the assumption that the monopole abundance was diluted by early \MMo~annihilations, and without taking into account the constraints coming from UV instantons and the early period of matter domination induced by the monopoles.
Equation~(\ref{eq:T_oscu}) then turns into
\beq
\!\!\!\!\!\!\!
    \frac{T_a}{10~\TeV} \approx \bl( \frac{p_0}{0.1} \br) \!\! \bl( \frac{266.25}{g_{*M}} \br) \!\! \bl( \frac{\alpha'}{0.02} \br) \!\! \bl( \frac{E'/N}{6} \br)^2  \!\! \bl( \frac{2.2 \times 10^{14}~\GeV}{f_a} \br)^2 \!\! \bl( \frac{1}{\xi_M T_M} \br)^3 \!\! \bl( \frac{1}{r_a T_S} \br) \!\! \bl( \frac{T_S/T_M}{4.6 \times 10^{-3}} \br) \!\! \bl( \frac{T_M}{10~\TeV} \br)  . \label{eq:stabilized_T_oscu}
\eeq

Thus, for the stabilized version of our model \Eqs{eq:fa_baseline_1}{eq:fa_baseline_2} become instead
\bea\label{eq:fa_stabilized_1}
    \text{for }T_a < T_M : \ \frac{f_a}{2.2 \times 10^{14}~\GeV} & \approx & \bl( \sqrt{\frac{\Omega_a h^2}{0.12}} \frac{2.15}{\theta_\infl} \br)^{0.48} \bl( \frac{p_0}{0.1} \frac{266.25}{g_{*M}} \br)^{0.36} \bl( \frac{\alpha'}{0.02} \br)^{0.36} \bl( \frac{E'/N}{6} \br)^{0.72} \nonumber\\
    && \times \bl( \frac{1}{T_M \xi_M} \br)^{1.1} \bl( \frac{1}{r_a T_S} \br)^{0.36} \ , \\
    \text{for } T_a > T_M: \ \frac{f_a}{2.2 \times 10^{14}~\GeV} & \approx & \bl( \sqrt{\frac{\Omega_a h^2}{0.12}} \frac{2.15}{\theta_\infl} \br)^{1.7} \bl( \frac{4.6 \times 10^{-3}}{T_S/T_M} \br)^{1.3} \ ,\label{eq:fa_stabilized_2}
\eea
where we have used $r_a T_S \approx 1$ as a benchmark.

\begin{figure}[t!]
\centering
\includegraphics[width=0.49\linewidth]{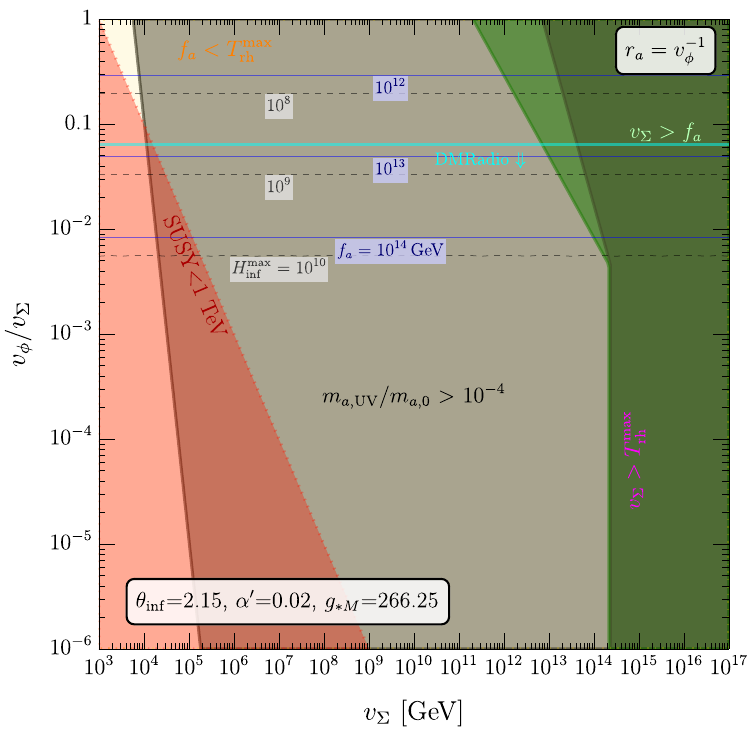}
\includegraphics[width=0.49\linewidth]{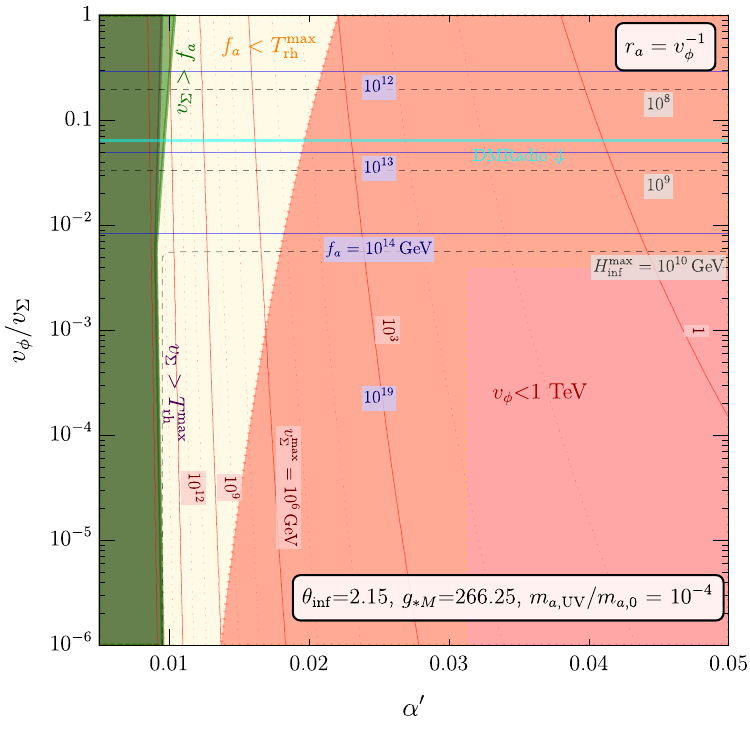}
\caption{Same as \Fig{fig:baseline}, except for a stabilized SUSY version of our model (and thus $g_{*M} = 266.25$).
In this model $r_a$ is the screening length associated with the Compton wavelength of the lightest charged fermion, the hidden higgsino $\tilde{\phi}$.
We then take $r_a T_S = m_{\tilde{\phi}}^{-1} T_S \approx 1$.
The left panel shows the $v_\Sigma$--$v_\phi/v_\Sigma$ parameter space (with $\alpha' = 0.02$), while the right panel shows the $\alpha'$--$v_\phi/v_\Sigma$ parameter space (with $v_\Sigma$ fixed such that $m_{a,\mathrm{UV}} = 10^{-4} \, m_{a,0}$).
The green, purple, yellow, and gray regions, as well as the blue, gray dashed, cyan, and red lines, have the same meaning as their counterparts in \Fig{fig:baseline}.
In red is the region for which $v_\phi < 1~\TeV$, and, since $m_{\tilde{\phi}} = m_\phi \approx v_\phi$, supersymmetric partners are found below $1~\TeV$.
Note that in this model $f_a$ cannot be larger than $\sim 10^{14}~\GeV$ regardless of the choice of $v_\Sigma$ and $v_\phi/v_\Sigma$, as shown in \Eq{eq:fa_stabilized_1}, unless $e'$, $\lambda_\Sigma$, and $\lambda_\phi$ (which control $\alpha'$, $\xi_M T_M$, and $r_a T_S$) be tuned in just the right way.
}
\label{fig:stabilized}
\end{figure}

In \Fig{fig:stabilized} we show the parameter space for the stabilized version of our model, where SUSY protects the $\phi$ potential from radiative corrections, at the cost of introducing another Higgs doublet $\phi$ (as is the case in SUSY) as well as the $\tilde{\phi}_{L,R}$ higgsinos, which screen the Witten effect with $r_a = m_{\tilde{\phi}}^{-1} = m_\phi^{-1}$.
As in the previous subsection, we assume $\xi_M^{-1} \approx T_M \approx v_\Sigma$, $T_S \approx v_\phi$, and take $p_0 = 0.1$, $E'/N = 6$, and $\theta_\infl = 2.15$.
The SUSY version of our benchmark model has two $\SUp$ scalar doublets and their corresponding fermionic partners, which constitute one additional vector-like doublet.
Thus, $N_f = 1$ in \Eq{eq:uv_inst}, and $m_i / v_\Sigma \approx v_\phi / v_\Sigma$.
In addition to this, since $\Sigma$ is part of a supermultiplet, there is a second scalar adjoint.
Assuming the minimal supersymmetric standard model as a benchmark, and accounting for all the HS superpartners, the total number of degrees of freedom in the stabilized model is $g_{*M} = 266.25$.
We also take $r_a T_S = T_S/m_\phi \approx 1$, and once again assume $\epsilon$ is large enough so that $T_\finu \approx T_S$.
Similarly to \Fig{fig:baseline}, the blue and gray contours correspond to $f_a$ in \Eqs{eq:fa_stabilized_1}{eq:fa_stabilized_2} and $\Hinf^{\rm max}$ in \Eq{eq:iso_Hinfl}, respectively.
The green, purple, gray, and yellow regions have the same meaning as in \Fig{fig:baseline}.
Since, at least in the HS, SUSY must be a good symmetry at the $v_\phi$ scale in order to protect the $\phi$ potential, we conservatively demand $v_\phi > 1~\TeV$ in order to satisfy collider constraints on superpartners.
We shade in red those points in parameter space for which this condition is not satisfied.

Note that, since $v_\phi \ll v_\Sigma$ is necessary in order to get a sufficiently small dilution factor $D$ for given $\theta_\infl$ and $f_a$, we have $r_a \sim v_\phi^{-1} \gg v_\Sigma^{-1} \sim r_c$.
Therefore, in this stabilized model, the $\beta$ factor in the monopole-induced axion mass is suppressed with respect to the baseline version, and thus the Witten effect cannot dynamically relax the axion misalignment angle to very small values.
As a result, $f_a$ cannot be as large as in the previous case, and in fact it cannot be raised much above $10^{14}~\GeV$ for our choice of parameter benchmark values, as can be deduced from \Eqs{eq:fa_stabilized_1}{eq:fa_stabilized_2}.
Indeed, in this version of the model, axion DM can have, at most, $f_a$ values in the lower end of DMRadio's reach, with its upper portion and the entirety of CASPEr-electric's reach being unattainable.
One can in principle dial $e'$, $\lambda_\Sigma$, and $\lambda_\phi$ just right, in order to change $\alpha'$, $T_M \xi_M$, and $r_a T_S$ in \Eqs{eq:fa_stabilized_1}{eq:fa_stabilized_2}.
However, the small exponents in these equations indicate that dialing these couplings can only increase the allowed values of $f_a$ by a modest amount.

We have seen that, in this stabilized version of our benchmark model, the large hierarchy between $v_\phi$ and $\Sigma$, necessary to obtain small $D$, can only raise $f_a$ so much.
However, throughout this analysis we have assumed $\epsilon$ is large enough that $T_\finu \approx T_S$, \ie, the \MMoS~system disappears soon after the formation of strings.
If $\epsilon$ were to be much smaller, this is no longer true.
In this case the \MMo~annihilations may be sufficiently delayed so that the Witten effect has enough time to relax the misalignment angle.
This extra time means that $T_S$ need not be that much smaller than $T_M$, and thus $v_\phi$ is not too far below $v_\Sigma$ either.
As a consequence of this, the screening of the dyonic hidden electric field may not be so severe, and $f_a$ may be raised further.
This is the realistic scenario, to which we turn our attention next.

\subsection{Realistic model case}

Let us now focus on the third version of our benchmark model, in which $v_\phi$ is as small compared to $v_\Sigma$ as naturally possible.
For example, radiative corrections to the scalar $\phi$ mass involving loops of the heavy gauge bosons $W^{\prime \pm}$ with $\vev{\Sigma}$ insertions will scale roughly like $\alpha'^2 v_\Sigma^2$.
Alternatively, if there are other $\OO(1)$ couplings (\eg, $\lambda_\Sigma$), we expect $v_\phi^2$ and $v_\Sigma^2$ to differ at most by a loop factor, or $1/16\pi^2$.
Therefore, we take $v_\phi / v_\Sigma \approx 0.1$ as the going rate of this natural hierarchy.

Since $v_\phi$ is so close to $v_\Sigma$, $\Up$ is Higgsed shortly after the original $\SUp$, and thus the string-forming PT takes place shortly after the PT responsible for the creation of monopoles, \ie, $T_S \lesssim T_M$.
The time passed between these two PTs is typically not nearly long enough to relax the axion misalignment angle to the required value.
This, in addition to the fact that that the monopole charges are confined below $T_S$, suggests that a monopole-induced dynamical relaxation simply does not occur to the required level.
However, this is a hasty conclusion.
Indeed, at distances below the flux tube thickness $\delta \sim m_{\gamma'}^{-1}$ the magnetic field lines appear unconfined.
Because of this we expect that, in the presence of an axion, the Witten effect takes place at sufficiently short distances.
As a result, the monopole still turns into a dyon, although its electric field now has a Yukawa suppression $\sim e^{-m_{\gamma'} r}$ due to the fact that the hidden photon has gained a mass.
The monopole-induced axion mass in \Eq{eq:U_axion} in the broken-$\Up$ phase must therefore be amended as follows:
\bea
    m_{a,M}^2 & \equiv & \frac{2 \beta n_M}{f_a} \ , \label{eq:ma_realistic} \\
    \text{with} \quad \beta & \approx & \frac{(E'/N)^2 \alpha' e^{-2 m_{\gamma'} \delta}}{32 \pi^2 \delta f_a} \ , \label{eq:beta_realistic}
\eea
instead of \Eq{eq:beta_axion}.
We have taken into account both the Yukawa suppression of the dyonic electric field as well as the cutoff distance of the axion--dyon system, which we expect to be given roughly by the thickness of the magnetic flux.

In this version of our benchmark model, the monopole--axion interactions will also dampen the axion misalignment angle, but only as long as the strings survive.
In order to give the axion enough time to relax, the $\epsilon$ portal coupling between the hidden and SM sectors has to be sufficiently small.
The smallness of $\epsilon$ has an additional effect.
Since $T_S$ is not too different from $T_M$, $\epsilon \ll 1$ ensures that the lightest HS particle (either $A'$ or one of the $\phi$s) is a long-lived massive relic.
The consistency condition requiring that the axion oscillations start before the onset of matter domination must include this new contribution to the total matter density; \Eq{eq:fa_MD} must therefore be amended.

Assuming that the couplings amongst the $\gamma'$ and $\phi$ particles are large enough, their mutual annihilations and decays occur in thermal equilibrium, thereby conserving the total entropy of the HS.
Once the heavier states have dumped their entropy into the lightest particle (which will occur below $T_S$, when they transition from relativistic to non-relativistic states), the comoving number density of this lightest state becomes constant.
If this lightest particle was still relativistic when this HS entropy dump took place, its number density is unchanged after it becomes non-relativistic, and it is ultimately determined by the HS entropy $s_{\rm HS}$: $n_{\rm light} a^3 = \text{const.} \approx (45 \zeta(3) / 2 \pi^4) s_{\rm HS} a^3$.
This allows us to compare the number density of the surviving non-relativistic HS relic to that of the monopoles:
\beq
    \frac{n_{\rm light}}{n_M} = \frac{n_{\rm light}/s_{\rm HS}}{(n_M a^3)/(s_{\rm HS} a^3)} \approx \frac{\zeta(3)}{\pi^2} \frac{g'_{*M} T_M^3}{p_0 \, \xi_M^{-3}} \ ,
\eeq
where $g'_{*M} = 13 + 7 N_f$ denotes the relativistic degrees of freedom in the HS at the time of monopole formation, where $N_f$ is the number of additional, light, vector-like fermions in the $\mathbf{2}$ irrep of $\SUp$.
As discussed in \Subsec{subsec:nelson_barr}, these fermions help suppress the UV instanton contributions to the axion mass (see \Eq{eq:uv_inst}).
We take as reference $N_f = 3$ fermions (\ie, $g'_{*M} = 34$, $g_{*M} = 140.75$), with identical masses equal to $m_i = m_\phi = \sqrt{\lambda} v_\phi$.
Thus, the screening of the dyon's hidden electric field due to the Compton wavelength of these fermions is as mild as possible.
Since $3 H_\MD^2 \mPl^2 = \rho_{\rm matter} = \rho_M + \rho_{\rm light}$ is the Hubble expansion rate during matter domination, we can finally write the consistency condition for the onset of axion oscillations:
\beq\label{eq:osc_condition}
    m_{a,M}^2 = \frac{2 \beta n_M}{f_a} > 9 H_\MD^2 = \frac{3 m_M n_M}{\mPl^2} \bl( 1 + \frac{\rho_{\rm light}}{m_M n_M} \br) \approx \frac{3 m_M n_M}{\mPl^2} \bl( 1 + 16 \, \frac{m_{\rm light}}{m_M} \bl( \xi_M T_M \br)^3 \!\! \bl( \frac{g'_{*M}}{34} \br) \!\! \bl( \frac{0.1}{p_0} \br) \br) \ .
\eeq
One can combine the previous equation with \Eqs{eq:ma_realistic}{eq:beta_realistic} to find the following consistency condition on $f_a$:
\bea\label{eq:fa_MD2}
    f_a & < & f_a^{\rm max} \equiv \frac{\alpha' E'/N \, e^{-1}}{2 \sqrt{6} \pi} \sqrt{\frac{v_\phi}{v_\Sigma}} \frac{\mPl}{\sqrt{1+x}} \approx 2.2 \times 10^{15}~\GeV \, \bl( \frac{\alpha'}{0.02} \br) \bl( \frac{E'/N}{6} \br) \sqrt{\frac{v_\phi/v_\Sigma}{0.1}} {\sqrt{\frac{1}{1+x}}} \ , \\
    \text{where} \quad x & \equiv & 0.17 \, \bl( \frac{0.1}{p_0} \br) \bl( \frac{g'_{*M}}{34} \br) \bl( \frac{\alpha'}{0.02} \br) \bl( \frac{m_{\rm light}}{m_{\gamma'}} \br) \bl( \xi_M T_M \br)^3 \bl( \frac{v_\phi/v_\Sigma}{0.1} \br) \label{eq:x_def}
\eea
is the second term inside the parentheses of the right hand side of \Eq{eq:osc_condition}.
In these expressions we used $m_{\gamma'}$ as a benchmark for $m_{\rm light}$, and took $\delta = m_{\gamma'}^{-1} = (e' v_\phi)^{-1}$.

As long as $f_a$ satisfies \Eq{eq:fa_MD2}, the monopole-induced axion oscillations and subsequent relaxation will occur.
In fact, these oscillations may in principle take place {\it both} during the unbroken $\Up$ phase (when the Universe had a temperature between $T_M$ and $T_S$), and during the broken $\Up$ (below $T_S$), leading to two periods of monopole-induced relaxation of the axion angle.
The first period starts when $m_{a,M} = 3 H$, with $\beta$ given by \Eq{eq:beta_axion}, and $r_a = m_f^{-1} \approx 1/(\sqrt{\lambda_\phi} v_\phi)$ (due to the light $N_f$ fermions).
This takes place at a temperature
\bea
    T_{a,1} & \approx & \frac{5}{8 \pi^4} \frac{p_0 (E'/N)^2 \, \alpha' \sqrt{\lambda_\phi}}{g_{*M}} \frac{\mPl^2 \, v_\phi}{f_a^2 \, \bl( \xi_M T_M \br)^3} \nonumber\\
    & \approx & 15~\TeV \, \bl( \frac{p_0}{0.1} \br) \!\! \bl( \frac{140.75}{g_{*M}} \br) \!\! \bl( \frac{\alpha'}{0.02} \br) \!\! \bl( \frac{E'/N}{6} \br)^2 \!\! \bl( \frac{\lambda_\phi}{1} \br)^{1/2} \!\! \bl( \frac{1.1 \times 10^{15}~\GeV}{f_a} \br)^2 \!\! \bl( \frac{1}{\xi_M T_M} \br)^3 \!\! \bl( \frac{v_\phi/v_\Sigma}{0.1} \br) \!\! \bl( \frac{v_\Sigma}{10~\TeV} \br) \ ,\label{eq:realistic_Ta1}
\eea
or at $T_M$, whichever is smallest.
Obviously, if $T_{a,1} < T_S$, this first period of oscillations during which $\Up$ is unbroken never actually takes place.
The second period also starts when $m_{a,M} = 3 H$, but this time with $\beta$ given by \Eq{eq:beta_realistic}.
This occurs either at a temperature
\bea
    T_{a,2} & \approx & \frac{5}{4 e^2 \pi^{7/2}} \frac{p_0 (E'/N)^2 \, \alpha'^{3/2}}{g_{*M}} \frac{\mPl^2 \, v_\phi}{f_a^2 \, \bl( \xi_M T_M \br)^3} \nonumber\\
    & \approx & 1~\TeV \, \bl( \frac{p_0}{0.1} \br) \!\! \bl( \frac{140.75}{g_{*M}} \br) \!\! \bl( \frac{\alpha'}{0.02} \br)^{3/2} \!\! \bl( \frac{E'/N}{6} \br)^2 \!\! \bl( \frac{1.1 \times 10^{15}~\GeV}{f_a} \br)^2 \!\! \bl( \frac{1}{\xi_M T_M} \br)^3 \!\! \bl( \frac{v_\phi/v_\Sigma}{0.1} \br) \!\! \bl( \frac{v_\Sigma}{10~\TeV} \br) \ ,\label{eq:realistic_Ta2}
\eea
or at $T_S$, whichever is smallest.
Because of this, the dilution in \Eq{eq:D_def} is broken up into two factors corresponding to these two periods, namely $D = D_1 \cdot D_2$, where
\beq\label{eq:dilution_realistic}
    D_1 \equiv  \min\bl[ 1, \, \bl( \frac{T_S}{\min\bl[ T_M, \, T_{a,1} \br]} \br)^{3/4} \br] \ , \quad
    \text{and} \quad D_2 \equiv \bl( \frac{T_\finu}{\min\bl[ T_S, \, T_{a,2} \br]} \br)^{3/4} \ . 
\eeq
Note that the definition of $D_1$ ensures that the unbroken $\Up$ phase does not contribute to the dilution factor if $T_{a,1} < T_S$, as argued above.
Clearly, the most dilution we can get out of this first phase is $D_1 = (T_S / T_M)^{3/4} \approx 0.18 \, ( (T_S/T_M) / 0.1)^{3/4}$.

We now turn our attention to the end of the axion angle relaxation.
In this version of our benchmark model we assume \Eq{eq:dM_xiS} is such that $d_M > \xi_S$.
This means that the strings lose energy via particle production, with a rate approximately given by
\beq
    \Gamma_\pp \approx \frac{\epsilon^2 v_\phi \lambda_\phi}{8 \pi^{5/2} \alpha'^{3/2}} \approx 1.2 \times 10^{-15}~\GeV \, \lambda_\phi \bl( \frac{\epsilon}{6.8 \times 10^{-10}} \br)^2 \bl( \frac{0.02}{\alpha'} \br)^{3/2} \bl( \frac{v_\phi/v_\Sigma}{0.1} \br) \bl( \frac{v_\Sigma}{10~\TeV} \br) \ ,
\eeq
where we have used $\xi_S \approx m_\phi^{-1}$ and $\delta \approx m_{\gamma'}^{-1}$, which are good approximations well below $T_S$, when temperature corrections to these quantities vanish.
To find the time at which the string dissipates all of its energy, we must compare $\Gamma_\pp$ to the Hubble expansion rate, which below $T_S$ scales like matter domination.
Thus, by demanding $\Gamma_\pp = H_{\rm MD} = \sqrt{m_M n_M (1+x) / (3 \mPl^2)}$, we find
\bea
    T_\finu & \approx & \frac{3^{1/3}}{4 \pi^{11/6}} \frac{\epsilon^{4/3} \lambda_\phi^{2/3}}{\alpha'^{5/6}} \frac{\xi_M T_M}{\bl( p_0 (1+x) \br)^{1/3}} \bl( \frac{\mPl^2 v_\phi^2}{v_\Sigma} \br)^{1/3} \nonumber\\
    & \approx & 13~\GeV \, \lambda_\phi^{2/3} \, \bl( \frac{\epsilon}{6.8 \times 10^{-10}} \br)^{4/3} \!\! \bl( \frac{0.02}{\alpha'} \br)^{5/6} \!\! \bl( \frac{0.1}{p_0} \br) \!\! \bl( \frac{1}{1+x} \br)^{1/3} \!\! \bl( \xi_M T_M \br) \!\! \bl( \frac{v_\phi/v_\Sigma}{0.1} \br)^{2/3} \!\! \bl( \frac{v_\Sigma}{10~\TeV} \br)^{1/3} \ . \label{eq:realistic_Tend1}
\eea
Implicit in the equation above is the assumption that no entropy dump from the HS into the SM has taken place before $T_\finu$.
Such an entropy dump will eventually take place, once the lightest HS particle decays.
For example, if the lightest particle is the dark photon, then $\Gamma_{\gamma'} \sim \epsilon^2 \alpha_{\rm em} m_{\gamma'}$ is its decay rate, and $\Gamma_{\gamma'} / \Gamma_\pp \sim 10^{-3} \, (\alpha' / 0.02)^2 / \lambda_\phi$, and the assumption is warranted for our benchmark parameters.

From this we can obtain the values of $f_a$ for which our dynamical misalignment mechanism guarantees the correct axion DM abundance in the realistic model:
\bea\label{eq:fa_realistic_1}
    \text{for } T_{a,2} < T_S : \ \frac{f_a}{1.1 \times 10^{15}~\GeV} & \approx & \bl( \sqrt{\frac{\Omega_a h^2}{0.12}} \frac{2.15}{\theta_\infl} \frac{p_0}{0.1} \frac{1}{\sqrt{\lambda_\phi}} \frac{0.18}{D_1} \br)^{0.48} \bl( \frac{140.75}{g_{*M}} \br)^{0.36} \bl( 1+x \br)^{0.12} \bl( \frac{\alpha'}{0.02} \br)^{0.84} \bl( \frac{E'/N}{6} \br)^{0.72} \nonumber\\
    && \times \bl( \frac{6.8 \times 10^{-10}}{\epsilon} \br)^{0.48} \bl( \frac{v_\Sigma}{10~\TeV} \br)^{0.24} \bl( \frac{v_\phi / v_\Sigma}{0.1} \br)^{0.12} \bl( \frac{1}{T_M \xi_M} \br)^{1.4} \ , \\
    \text{for } T_{a,2} > T_S: \ \frac{f_a}{1.1 \times 10^{15}~\GeV} & \approx & \bl( \sqrt{\frac{\Omega_a h^2}{0.12}} \frac{2.15}{\theta_\infl} \frac{1}{\sqrt{\lambda_\phi}} \frac{0.18}{D_1} \br)^{1.7} \bl( \frac{p_0}{0.1} \br)^{0.43} \bl( 1+x \br)^{0.43} \bl( \frac{\alpha'}{0.02} \br)^{1.1} \nonumber\\
    && \times \bl( \frac{6.8 \times 10^{-10}}{\epsilon} \br)^{1.7} \bl( \frac{T_S^3 / v_\Sigma}{\bl( 1~\TeV \br)^3 / 10~\TeV} \br)^{0.43} \bl( \frac{0.1}{v_\phi/v_\Sigma} \br)^{0.86} \bl( \frac{1}{T_M \xi_M} \br)^{1.3} \ .
    \label{eq:fa_realistic_2}
\eea

\begin{figure}[t!]
\centering
\includegraphics[width=0.49\linewidth]{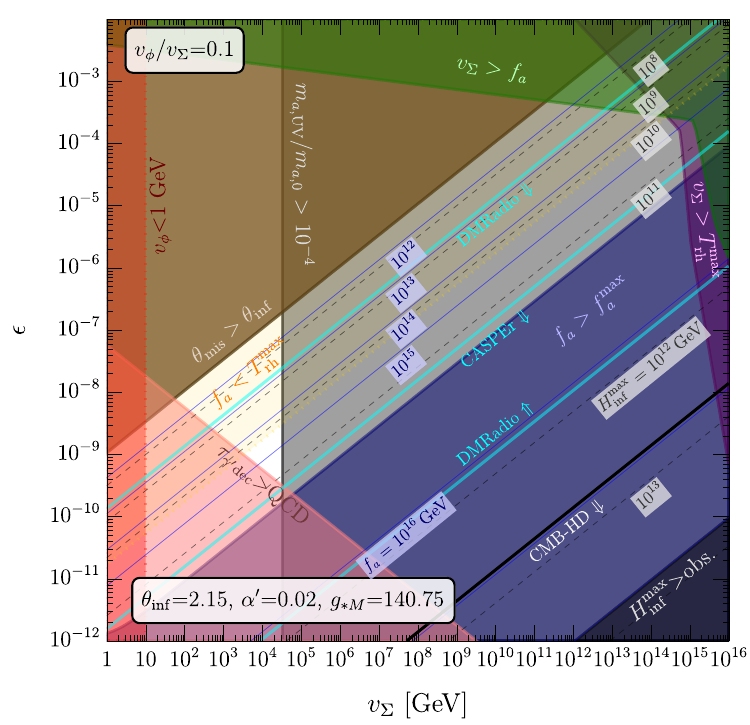}
\includegraphics[width=0.49\linewidth]{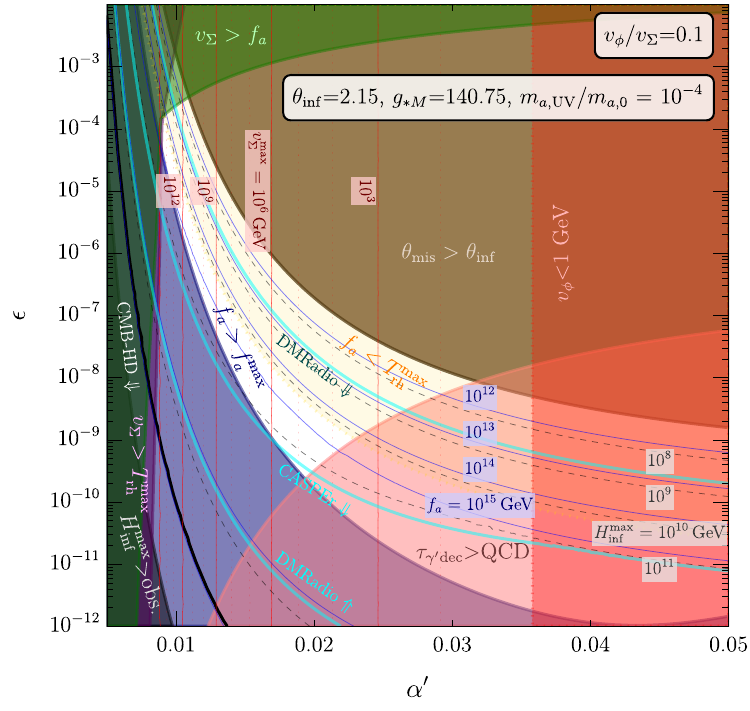}
\caption{Same as \Fig{fig:baseline}, except for the realistic version of our model; we assume $N_f = 3$ additional $\SUp$ doublet vector-like fermions, with masses $m_i / v_\Sigma \approx m_\phi / v_\Sigma$.
As a result, $g_{*M} = 140.75$
We also assume $\xi_M^{-1} \approx T_M \approx v_\Sigma$, $T_S \approx v_\phi$, $\xi_S \approx m_\phi^{-1} = 1/(\sqrt{\lambda_\phi} v_\phi)$, $r_a \approx \delta \approx m_{\gamma'}^{-1}$, $m_{\rm light} = m_{\gamma'}$.
We take $v_\phi/v_\Sigma = 0.1$, and $\lambda_\phi = 1$.
The blue, green, purple, red, yellow, and gray regions, as well as the blue, gray dashed, cyan, and red lines, have the same meaning as their counterparts in \Fig{fig:baseline}.
The black region shows where $\Hinf > 6.1 \times 10^{13}~\GeV$; the black line indicates the sensitivity of CMB-HD to the inflationary $r$-parameter and through it, to $\Hinf^{\rm max}$ and $f_a$.
The brown region excludes the nonsensical $\theta_\mis > \theta_\infl$ (\ie, $T_\finu > T_\oscu$), while the pink region excludes those points for which the lightest hidden sector state, here assumed to be $A'$, has a lifetime longer than the Hubble timescale at $1~\GeV$, roughly when the QCD phase transition occurred.
}
\label{fig:realistic}
\end{figure}

Figure~(\ref{fig:realistic}) shows the parameter space of the realistic version of our model.
As in the previous two versions, we take $p_0 = 0.1$, $E'/N = 6$, and $\theta_\infl = 2.15$.
We assume there are $N_f = 3$ additional vector-like fermion $\SUp$ doublets mitigating the contribution to the axion mass from UV instantons, which means $g_{*M} = 140.75$.
The mass of these extra fermions is taken to be roughly equal to $m_\phi$, and thus $m_i/v_\Sigma \approx \sqrt{\lambda_\phi} v_\phi / v_\Sigma \approx \sqrt{\lambda_\phi} \, 0.1$.
In addition, we take $\xi_M^{-1} \approx T_M \approx v_\Sigma$, $T_S \approx v_\phi$, $\xi_S \approx m_\phi^{-1} = 1/(\sqrt{\lambda_\phi} v_\phi)$, and $r_a \approx \delta \approx m_{\gamma'}^{-1}$.
We assume $\lambda_\phi = 1$.
We further assume that the lightest hidden sector particle is the hidden photon, and therefore take $m_{\rm light} = m_{\gamma'}$ in \Eq{eq:x_def}.
In blue we show the axion DM $f_a$ contours obeying \Eqs{eq:fa_realistic_1}{eq:fa_realistic_2}, while in gray the $\Hinf^{\rm max}$ contours that satisfy \Eq{eq:iso_Hinfl}.
The cyan contours indicate the sensitivities of DMRadio and CASPEr-electric, while the thick black line indicates the value of $\Hinf^{\rm max}$ whose corresponding tensor-to-scalar ratio $r$ can be detected by CMB-HD.
As in \Fig{fig:baseline}, the blue, purple, green, yellow, and red regions represent the various constraints and consistency conditions necessary for the dynamical axion misalignment mechanism induced by the Witten effect to take place.
Inside the black region are those points for which $\Hinf^{\rm max}$ is above the Planck+BICEP2/Keck Array constraint of $6.1 \times 10^{13}~\GeV$.
In brown and pink we include two other additional constraints.
The brown region excludes those points for which dynamical relaxation of the axion angle is unnecessary.
Indeed, this region corresponds to $T_\finu > T_\oscu$ or, equivalently, $D > 1$ and $\theta_\mis > \theta_\infl$, which is of course nonsensical.
The pink region shows those points for which the lightest hidden sector particle, in this case the dark photon, has a lifetime ($\tau_{\gamma'} \sim (\epsilon^2 \alpha_{\rm em} m_{\gamma'})^{-1}$) longer than the age of the Universe at temperatures of $1~\GeV$, roughly the time of the QCD phase transition ($t \sim 1 \times 10^{-18}~\GeV^{-1}$).
We have conservatively excluded this region, since for the parameter points within it the era of early matter domination caused by the relic $A'$ particles extends beyond the QCD phase transition.
This means that the $A'$ dump their entropy too late, and thus can upset the standard axion misalignment mechanism.
As shown in this figure, the realistic version of our benchmark model can raise $f_a$ above $10^{15}~\GeV$.
This is an order of magnitude better than the supersymmetric stabilized version, although it is still insufficient to allow $\Hinf$ values safe from isocurvature bounds with a corresponding $r$-parameter observable by CMB-HD and other future CMB experiments.
However, a good portion of DMRadio's sensitivity is recovered compared to the stabilized scenario, as well as part of CASPEr-electric's.
Thus, a realistic hidden sector with $v_\Sigma \sim 10^5~\GeV$ and $\alpha' \approx 0.02$ (and thus a monopole mass $m_M \sim 10^6~\TeV$) can reach the largest values of $f_a$, as long as $\epsilon \sim 10^{-10}$.

We conclude this section by noting that one could, in principle, combine elements of the stabilized and realistic models. For example, a hierarchy of VEVs ($v_\phi/v_\Sigma \ll 1$) stabilized by SUSY or another mechanism, alongside a small $\epsilon$, can naturally lead to a separation of temperature scales $T_\finu \ll T_S \ll T_M$. Following \Eq{eq:dilution_realistic}, this would allow for greater dilution of the misalignment angle in the $\Up$ unbroken phase. Stabilization mechanisms introduce light fermions with masses $\sim v_\phi$ (as noted near \Eq{eq:stabilized_T_oscu}). These fermions also suppress UV instanton contributions to the axion mass, eliminating the need for ad-hoc fermions. Thus, combining both models could be highly effective. However, we refrain from delving further, as we have already outlined the necessary components.

\section{Conclusions}
\label{sec:conclusions}

In this paper we constructed a model in which a pre-inflationary QCD axion can constitute the entirety of the dark matter, have a large decay constant $f_a$ (or, equivalently, a small mass), and avoid isocurvature constraints, all without tuning the initial axion misalignment angle $\theta_\infl$ coming out of inflation.
This is achieved with the help of a population of magnetic monopoles of a hidden sector coming into existence in the early Universe.
These monopoles provide the axion with a large yet temporary mass via the Witten effect.
This mass contribution, being proportional to the hidden monopole abundance, evolves in time as the Universe expands.
Provided the axion-monopole coupling is strong enough, this will result in the onset of axion field oscillations much earlier than in the usual case of the QCD axion misalignment mechanism.
This earlier phase of oscillations relaxes the axion misalignment angle, reducing it from its generic initial value right after inflation to a much smaller $\theta_\mis$.
The subsequent Higgsing of the hidden electromagnetic symmetry at a lower scale results in the confining and eventual annihilation of these monopole magnetic charges through the formation of strings of magnetic flux tubes.
Such a process is necessary to ensure that the energy density of the Universe is not monopole-dominated, and for the monopole-induced axion mass to go away, leaving the axion mass to be dominated by its QCD contribution at low energies.
After these monopoles have disappeared, a second era of axion oscillations, corresponding to those of the misalignment mechanism of the standard QCD axion, proceeds as normal but with smaller initial angle $\theta_\mis$, thereby allowing the axion to reproduce the known dark matter abundance.

The kernel of this idea can be found in various forms in previous literature (\eg, Refs.~\cite{Kawasaki:2015lpf,Nomura:2015xil,Kawasaki:2017xwt}).
In our present work we map out, for the first time, all the necessary conditions that a given model must satisfy in order for this monopole-induced dynamical relaxation of the axion misalignment angle to take place.
By working out the details of one specific benchmark model, that of $\SUp \to \Up$ spontaneous symmetry breaking, we showed that ({\it i.}) the hidden monopole abundance can be significantly larger than previously thought; that ({\it ii.}) the onset of axion oscillations is severely limited by a robust upper-limit on $f_a$; that ({\it iii.}) the parameter space is further constrained by requiring that the irreducible UV instanton contribution to the axion mass leave the QCD axion solution to the strong CP problem intact; and that ({\it iv.}) the hitherto unexplored possibility of having the monopoles confine soon after their formation does not entirely hinder the dynamical relaxation of the axion angle, provided that the portal coupling between the hidden and standard model sectors be small enough.
In addition, by including a very mild Nelson-Barr mechanism as part of our UV model, we can ensure that the minima of the monopole-induced and QCD axion potentials be sufficiently aligned for the whole idea to work, without invoking a coincidence or anthropic tuning between them.
We studied three variants of our general model, each of them distinguished by how the scale of $\Up$ breaking compares to the larger scale of $\SUp$ breaking: either ({\it a}) fine-tuned (the ``baseline'' case), ({\it b}) protected from radiative corrections by supersymmetry (the ``stabilized'' case), or ({\it c}) as small radiative corrections permit it to be (the ``realistic'' case).

In all cases we found that the consistency conditions outlined above severely constrain the $\Up$ fine structure coupling $\alpha'$.
More concretely, we find that $0.01 \lesssim \alpha' \lesssim 0.03$: smaller values are inconsistent with the pre-inflationary axion picture in question, while larger values result in a prohibitively small $\Up$ breaking scale.
In general, we find that the scale of $\SUp$ breaking, which is related to the monopole mass, can be as large as $\sim 10^{15}~\GeV$ and as low as $10~\GeV$, with $1$--$100~\TeV$ typically leading to the smallest $\theta_\mis$, and consequently the largest $f_a$ scales.
Indeed, we found that $f_a$ can be several orders of magnitude larger than the $10^{12}~\GeV$ predicted by the standard QCD misalignment mechanism, easily reaching $10^{14}~\GeV$ and even $10^{16}~\GeV$.
As mentioned in the introduction, large-$f_a$ QCD axion dark matter is the target of experiments such as DMRadio and CASPEr-electric, which are unable to probe the parameter space of the untuned, standard picture.
The pre-inflationary axion scenario is safe from isocurvature constraints as long as the Hubble scale during inflation, $\Hinf$, is sufficiently low.
By raising $f_a$, our model allows a correspondingly larger $\Hinf$ (up to $10^{11}~\GeV$); however, the inflationary scale remains too low for primordial gravitational waves to be observable by upcoming experiments such as CMB-HD and LiteBIRD.
Nevertheless, we expect that parts of our parameter space can be probed in different experiments.
Indeed, since the $\SUp$ and $\Up$ breaking scales can be as low as $\TeV$ and $\GeV$, respectively, the resulting massive hidden photons could be detected in future collider or beam dump experiments, if their kinetic mixing with the SM photon is large enough.

We trust that our current work has demonstrated how axion and monopole physics could play an important role in the cosmological history of the Universe, relax the three-way tension between the high-scale pre-inflationary axion, axion dark matter, and isocurvature, and at the same time motivate the target parameter space of state-of-the-art axion dark matter experiments.

\acknowledgments{
The authors would like to thank Anson Hook for his invaluable contributions in the early stages of this work.
We would also like to thank Zackaria Chacko and Raman Sundrum for useful discussions.
MBA thanks Stephanie Buen Abad for proofreading this manuscript.
AB and MBA supported by NSF grant PHY-2210361 and the Maryland Center for Fundamental Physics.}

\bibliographystyle{utphys}
\bibliography{reference.bib}

\end{document}